\documentclass[reprint, amsmath,amssymb,aps,pre]{revtex4-2}

\usepackage[pdfborder={0 0 0.5 [3 2]}, plainpages=false]{hyperref}%
\usepackage[left=1in,right=1in,top=1in,bottom=1in]{geometry}%

\usepackage{amsmath}
\usepackage{enumerate}
\usepackage{amssymb}                
\usepackage{amsfonts}
\usepackage{amsthm}
\usepackage{bbm}
\usepackage[table,xcdraw]{xcolor}
\usepackage{mathtools}
\usepackage{cool}
\usepackage{graphicx,epsfig}

\usepackage{subcaption}
\usepackage[capitalize,nameinlink]{cleveref}
\crefname{section}{section}{sections}
\crefname{subsection}{subsection}{subsections}
\Crefname{section}{Section}{Sections}
\Crefname{subsection}{Subsection}{Subsections}

\Crefname{figure}{Figure}{Figures}

\crefformat{equation}{\textup{#2(#1)#3}}
\crefrangeformat{equation}{\textup{#3(#1)#4--#5(#2)#6}}
\crefmultiformat{equation}{\textup{#2(#1)#3}}{ and \textup{#2(#1)#3}}
{, \textup{#2(#1)#3}}{, and \textup{#2(#1)#3}}
\crefrangemultiformat{equation}{\textup{#3(#1)#4--#5(#2)#6}}%
{ and \textup{#3(#1)#4--#5(#2)#6}}{, \textup{#3(#1)#4--#5(#2)#6}}{, and \textup{#3(#1)#4--#5(#2)#6}}

\Crefformat{equation}{#2Equation~\textup{(#1)}#3}
\Crefrangeformat{equation}{Equations~\textup{#3(#1)#4--#5(#2)#6}}
\Crefmultiformat{equation}{Equations~\textup{#2(#1)#3}}{ and \textup{#2(#1)#3}}
{, \textup{#2(#1)#3}}{, and \textup{#2(#1)#3}}
\Crefrangemultiformat{equation}{Equations~\textup{#3(#1)#4--#5(#2)#6}}%
{ and \textup{#3(#1)#4--#5(#2)#6}}{, \textup{#3(#1)#4--#5(#2)#6}}{, and \textup{#3(#1)#4--#5(#2)#6}}

\crefdefaultlabelformat{#2\textup{#1}#3}

\hypersetup{hidelinks}

\def\Z{{\mathbb Z}}

\graphicspath{ {images/} }

\begin{document}

\title{Floquet solitons in square lattices: Existence, Stability and Dynamics}
% \author{Ross Parker}
% \date{June 2021}

\author{Ross Parker}
\affiliation{Department of Mathematics, Southern Methodist University, 
Dallas, TX 75275, USA}
\email{rhparker@smu.edu}

\author{Jes\'us Cuevas-Maraver}
\affiliation{Grupo de F\'{\i}sica No Lineal, Departamento de F\'{\i}sica Aplicada I,
Universidad de Sevilla. Escuela Polit\'{e}cnica Superior, C/ Virgen de Africa, 7, 41011-Sevilla, Spain}
\affiliation{Instituto de Matem\'{a}ticas de la Universidad de Sevilla (IMUS). Edificio
Celestino Mutis. Avda. Reina Mercedes s/n, 41012-Sevilla, Spain}

\author{P.\,G. Kevrekidis} 
\affiliation{Department of Mathematics and Statistics, University of Massachusetts, Amherst MA 01003, USA}
\email{kevrekid@math.umass.edu}

\author{Alejandro Aceves}
\affiliation{Department of Mathematics, Southern Methodist University, 
Dallas, TX 75275, USA}
\email{aaceves@smu.edu}

\begin{abstract}
In the present work, we revisit a recently proposed and experimentally realized topological 2D lattice with periodically time-dependent
interactions. We identify the fundamental solitons, previously
observed in experiments and direct numerical simulations, as
{\it exact}, exponentially localized, periodic in time solutions.
This is done for a variety of phase-shift angles of the central
nodes upon an oscillation period of the coupling strength.
Subsequently, we perform a systematic Floquet stability analysis
of the relevant structures. We analyze both their point
and their continuous spectrum and find that the solutions are
generically stable, aside from the possible emergence
of complex quartets due to the collision of bands of 
continuous spectrum. The relevant instabilities become
weaker as the lattice size gets larger. Finally,
we also consider {\it multi-soliton} analogues of these
Floquet states, inspired by the corresponding discrete
nonlinear Schr{\"o}dinger (DNLS) lattice. When exciting 
initially multiple sites in phase, we find that the
solutions reflect the instability of their DNLS
multi-soliton counterparts, while for configurations
with multiple excited sites in alternating phases,
the Floquet states are spectrally stable, again analogously to their DNLS counterparts.
\end{abstract}

\maketitle

\section{Introduction}

The study of topological features and their interplay with the
dynamics is a theme of growing significance in a diverse variety
of fields including photonics \cite{Ozawa2019}, cold
atom physics~ \cite{Cooper2019}, as well as phononics~\cite{Ma2019,Susstrunk2016} and metamaterials~\cite{Bertoldi}, among others. 
While much of the relevant emphasis has been on linear
features of relevant models, progressively there is an increasing
number of studies at the interface between 
nonlinearity and topology~\cite{Smirnova2020,Ma2021}; see also the
corresponding chapter of~\cite{saxena}. 

In the context of nonlinear systems, there has been progress
in a number of pertinent directions. For instance, nonlinearity
has been leveraged in order to modulate the frequency and generate the harmonics of edge states~\cite{Dobrykh2018, Pal2018, Vila2019, Kruk2019, Wang2019, Darabi2019, Zhou2020}. Moreover, coherent nonlinear
wave structures that are dynamically robust and potentially 
propagate on edges of domains in the context of models with
suitable topology have been identified~\cite{Ablowitz2014, Leykam2016, Kartashov2016, Snee2019, Tao2020}. Among the numerous 
further states that have been explored, one can mention
nonlinear Dirac cones \cite{Bomantara2017}, gap solitons
induced by topological bands~\cite{Lumer2013, Solnyshkov2017, Smirnova2019, Marzuola2019, Mukherjee2020}, as well as domain walls~\cite{Chen2014, Hadad2017, Poddubny2018}. Features such as the uninhibited unidirectional,
scatter-free (around lattice defects) propagation of nonlinear
edge modes in topological lattices 
(such as Lieb, Kagom{\'e} etc.)~\cite{Abl19a}, as well as the absence
of Peierls-Nabarro, discreteness-induced barriers in nonlinear
Floquet topological insulators~\cite{Abl21a} have been manifested.
These suggest the particular promise of topological nonlinear media
in overcoming some of the limitations of ``conventional'' nonlinear
modes. Recently, relevant topological phase transitions have been
extended to entire soliton lattices~\cite{chen21a}.

In the present work, our aim is to explore systematically
a model of an anomalous Floquet topological insulator that
has not only been recently proposed, but also experimentally
implemented in~\cite{Mukherjee2020}. In the relevant context,
a periodically modulated waveguide lattice was produced, with 
the Floquet (periodic) driving inducing a nonvanishing 
winding number. This, in turn, was argued to produce topological
edge modes in the relevant spectrum. The topological bandgap
produced in such a medium, in the presence of cubic nonlinearity
due to the optical Kerr effect, was found to lead to the formation 
of solitary waves that could be experimentally observed in~\cite{Mukherjee2020}. While such waves were identified and
the extent of their spatial localization was examined for 
different input powers in this work, an understanding of such
states is still rather limited from the nonlinear and dynamical
systems point of view. 

Here, we offer a systematic exploration of the existence and
stability of such states. Imposing a Floquet, time-periodic
modulation of the waveguide coupling emulating that of the
experiment, we seek and are able to identify such states as
{\it numerically exact} solutions, up to a prescribed 
numerical accuracy. This is done for a variety of phase-shifts
arising for each period of the coupling modulation (such as, e.g.,
$\pi$, $\pi/2$, etc.), extending in this
way the direct simulations and experiments of~\cite{Mukherjee2020}. 
Once the relevant
waveforms have been identified, a natural subsequent question
is that of their dynamical stability. The experimental observation
of such states predisposes towards their stability and hence observability,
yet the parametric range of such a feature is of particular interest.
Indeed, here we report a systematic Floquet analysis which reveals that the relevant fundamental states are {\it spectrally stable}, featuring
multipliers purely on the unit circle for  wide parametric intervals. 
Interestingly, the relevant spectrum is found to consist of a continuous
spectrum surrounding $(1,0)$ in the Floquet multiplier plane, and of
a few point-spectrum multipliers associated with the excitations
of the core of the relevant solitary wave. Instabilities emerge
as a byproduct of the finite size of the numerical computations
and have been dynamically monitored, yet relevant features of the
continuous spectrum are found to weaken for progressively larger
lattices and hence are expected to be absent in the infinite
lattice limit. Finally, another question that stems from a 
well-rounded understanding of the corresponding non-time-modulated
analogue of the model, namely the discrete nonlinear
Schr{\"o}dinger (DNLS) equation~\cite{kev09}, is whether additional
coherent structures may exist in such a setting. Indeed, here we
illustrate a systematic prescription to produce multi-soliton
states, which is motivated from the corresponding multi-soliton
states of the DNLS. In particular, we show that an initial excitation of multiple sites 
(adjacent or diagonally) produces a corresponding multi-site 
Floquet topological soliton. Furthermore, the stability structure
of the DNLS is found to carry over to the temporally modulated coupling model:
in-phase excited sites are associated with a real pair of Floquet 
multipliers. On the other hand, out-of-phase excited states, the so-called twisted modes, are associated with
a stable Floquet spectrum and long-lived multi-soliton periodic orbits.
All of the above results have been corroborated by systematic numerical simulations. 

The presentation of our results is structured as follows. 
In section 2, we discuss the model and the specific choices
of initial and boundary conditions, as well as the concrete
temporal modulation of the coupling. In section 3, we explore
the fundamental breather states of the lattice inspired by
(and substantially extending the results of)~\cite{Mukherjee2020}.
In section 4, we leverage the detailed understanding of the
DNLS model to extend considerations to multi-peak Floquet
solitons and to examine their stability properties. Finally,
in section 5, we summarize our findings and present some
directions for future study. 

\section{Mathematical model}

\begin{figure}
    \centering
    \begin{subfigure}{0.49\linewidth}
    \caption{}
    \label{fig:modelcartoon}
    \includegraphics[width=4.1cm]{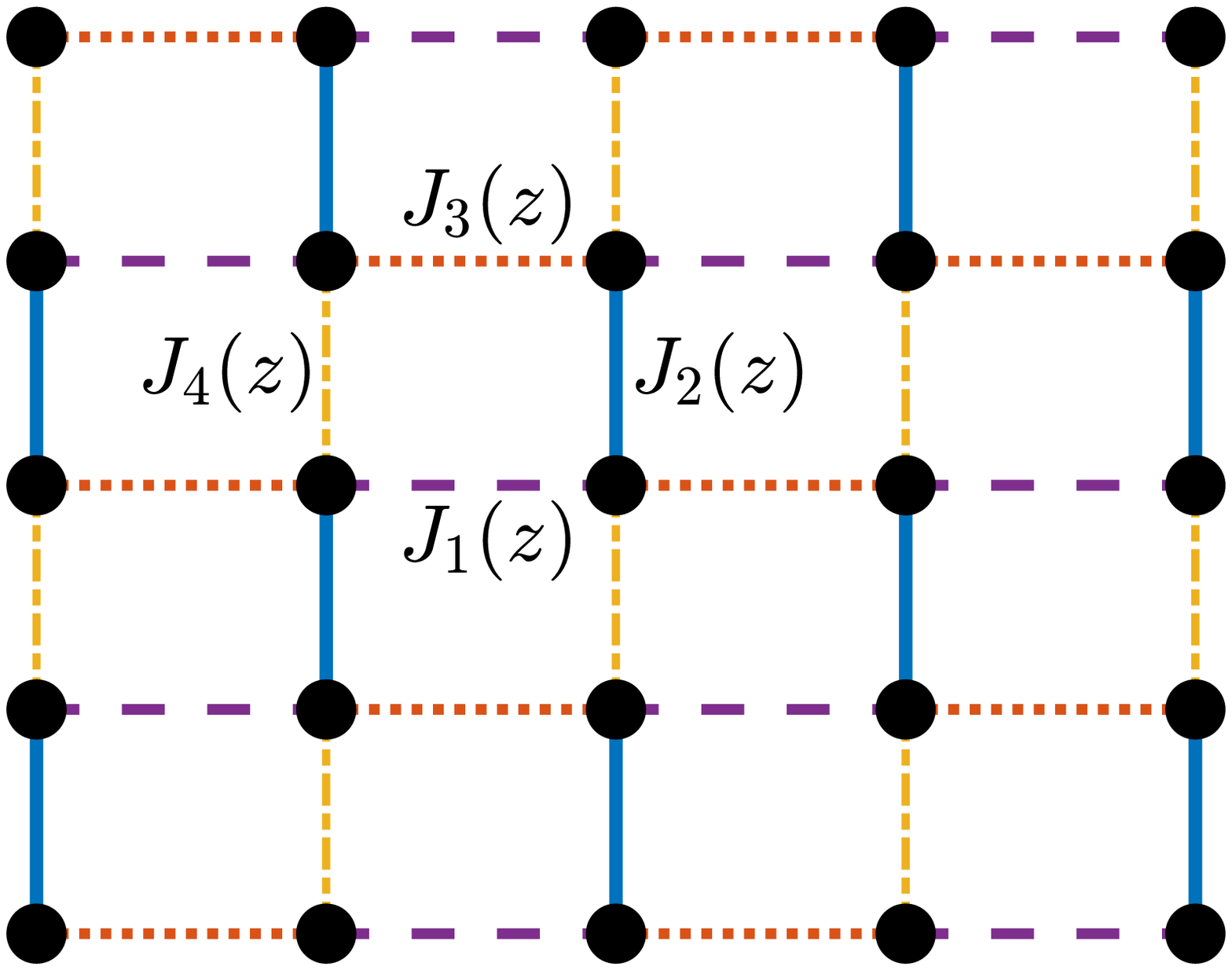}
    \end{subfigure}
    \begin{subfigure}{0.49\linewidth}
    \caption{}
    \includegraphics[width=4.1cm]{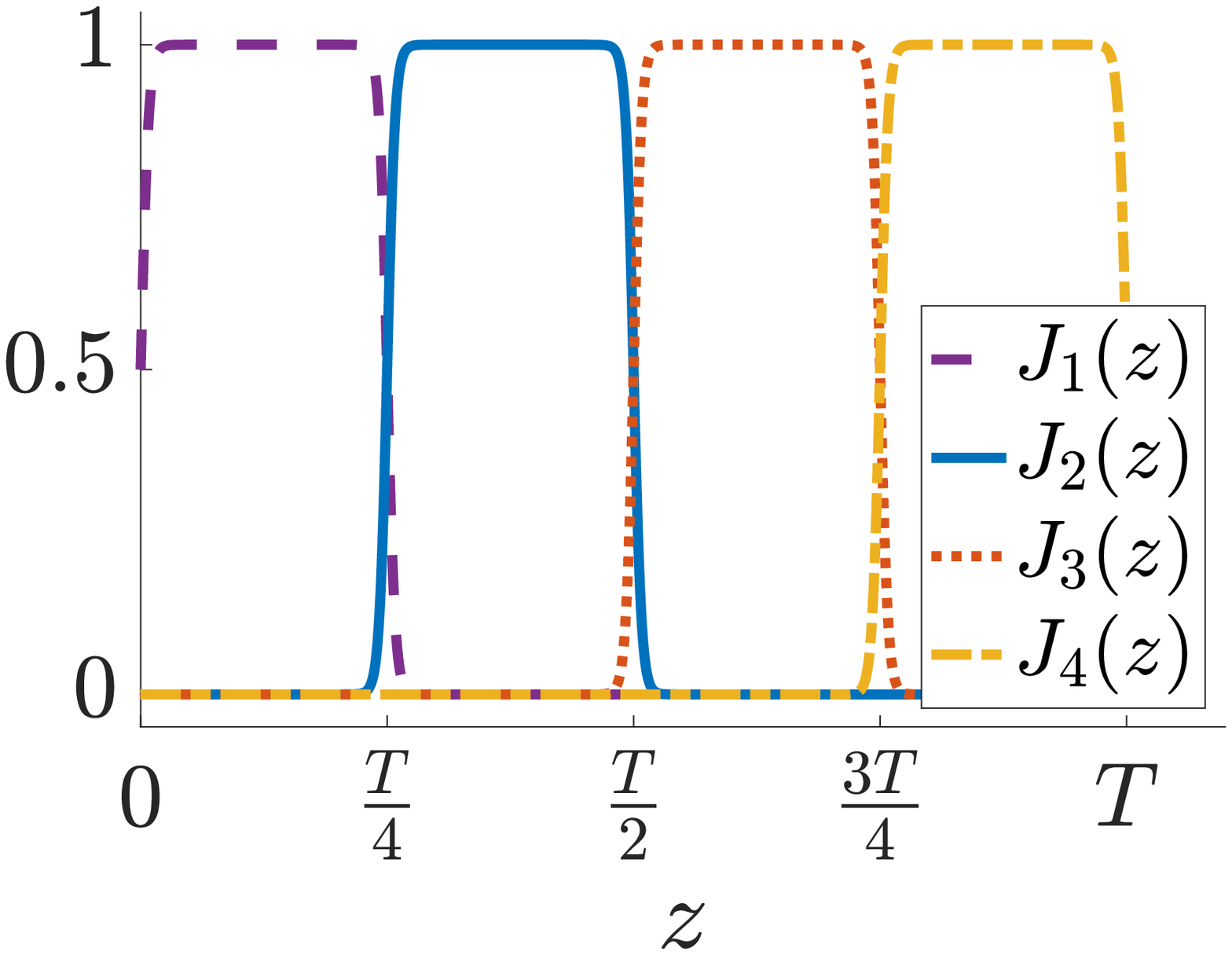}
    \end{subfigure}
    \caption{(a) Cartoon of $z$-dependent nearest-neighbor couplings. (b) Coupling functions $J_j(z)$ over one period of the coupling for the choice
    of the period $T = 2 \pi$ with steepness parameter $k=20$.}
    \label{fig:model}
\end{figure}

The propagation of light through an optical lattice with nearest-neighbor coupling in the presence of a third-order Kerr nonlinearity can be described by the discrete nonlinear Schr\"odinger equation 
\begin{equation}\label{eq:model}
i \frac{d}{dz} \phi_s(z) = C \sum_{\langle s' \rangle} H_{ss'}(z) \phi_{s'} - \gamma |\phi_s|^2 \phi_s,
\end{equation}
where $C$ is the coupling strength, $\gamma$ is the strength of the nonlinearity, which we will always take to be 1.
This non-dimensional version represents experimental conditions for nonlinear modes of milliwatts peak power and millimeter effective nonlinear length (that is, the non-dimensional intensity $|\phi_s|^2 = O(mm^{-1})$) ~\cite{Mukherjee2020}. $H_{ss'}(z)$ is the linear tight-binding, nearest-neighbor coupling, which depends on the propagation distance $z$~\cite{Mukherjee2020} (and has an explicit form shown in Eqs.~\ref{eq:model2} below). The summation is over nearest neighbors only. We consider here a square lattice of waveguides in which the strengths of the nearest-neighbor couplings vary periodically in $z$ with fundamental period $T$ in such a way that for each $z$, every waveguide only interacts with one of its four neighbors (\cref{fig:modelcartoon}). For the two-dimensional integer lattice $\Z^2$, equation \cref{eq:model} becomes
\begin{equation}\label{eq:model2}
\begin{aligned}
i \frac{d}{dz} &\phi_{m,n}(z) = \\
&\begin{cases}
C \Big( J_1(z)\phi_{m+1,n}(z) + J_2(z)\phi_{m,n-1}(z) \\
    \qquad  J_3(z)\phi_{m-1,n}(z) + J_4(z)\phi_{m,n+1}(z) \Big) \\
    \qquad -\gamma |\phi_{m,n}(z) |^2 \phi_{m,n}(z) \qquad\quad m+n \textrm{ even} \\
C \Big( J_1(z)\phi_{m-1,n}(z) + J_2(z)\phi_{m,n+1}(z) \\
    \qquad + J_3(z)\phi_{m+1,n}(z) + J_4(z)\phi_{m,n-1}(z) \Big) \\
    \qquad -\gamma |\phi_{m,n}(z) |^2 \phi_{m,n}(z) \qquad\quad m+n \textrm{ odd}.
\end{cases}
\end{aligned}
\end{equation}
The functions $J_j(z)$ model the switching of neighbor coupling as follows. For period $T$, we define the smoothed bump function
\begin{equation}
J(z) = \frac{1}{2} \left[ \tanh(k z) - \tanh\left(k \left(z-\frac{T}{4}\right)\right)\right],
\end{equation} 
where the parameter $k$ quantifies the steepness of the bump, with steepness increasing with $k$. The four coupling functions in \cref{fig:model} are then given by $J_1(z) = J(z)$, $J_2(z) = J(z-T/4)$, $J_3(z) = J(z - T/2)$, and $J_4(z) = J(z-3T/4)$. Notice that effectively on any propagation distance interval of length $T/4$, there is only one active coupling in place (see Figure 1b).

\section{Fundamental breather}

The system \cref{eq:model} has a fundamental breather solution in which the optical intensity is localized in a square of four lattice sites (which we call the fundamental unit square) and rotates counterclockwise around these sites (\cref{fig:fundbreather}(a) and (b)). After one period $T$, the solution reproduces itself except for a phase shift $\theta$. Numerical simulations demonstrate that this solution can be found with phase shifts of $\theta = \pi$, $\pi/2$, $\pi/3$, and $\pi/4$, but $\theta = 0$ is not possible. (Numerical simulations suggest that solutions for $\theta = \pi/n$ for integers $n > 4$ can be obtained, but this was practically found to be increasingly more difficult to do using a shooting method for larger $n$.)
In line with the original work of~\cite{Mukherjee2020},
the phase $\theta=0$ is associated with a quasi-energy
resonant with the linear
band of extended excitations and hence is not possible. The overall period of the breather is then given by $\tau = m T$, where $m = \frac{2\pi}{\theta}$ is the number of periods $T$ needed to return to the starting condition. We will consider herein only the phase shifts $\theta = \pi$ and $\pi/2$, but will briefly comment on what occurs in the other cases.

We use a shooting method to construct a breather solution numerically. We first choose the period $T$ and the phase shift $\theta$. For all of our simulations, we will use a period $T = 2 \pi$ of the coupling time-dependence. If $\theta = \pi$, for example, the overall period of the breather will be $\tau = 4 \pi$. Starting with an initial guess $\phi(0)$, we evolve the solution forward using a 4th order Runge-Kutta scheme. To obtain a solution with period $T$ and phase shift $\theta$, we iteratively solve $\phi(0) e^{i \theta} - \phi(T) = 0$. We also compute the Floquet spectrum of the breather solution by determining the eigenvalues of the monodromy matrix over one full breather period $\tau$. For efficiency of computation, we run the simulation on a $20\times 20$ lattice with periodic boundary conditions. For the fundamental breather, a single lattice site is excited for the initial guess.

\subsection{Phase shift \texorpdfstring{$\theta = \pi$}{pi}}\label{sec:phasepi}

\begin{figure}
    \centering
    \begin{subfigure}{0.49\linewidth}
    \caption{}
    \label{fig:fundbreathera}
    \includegraphics[width=4.1cm]{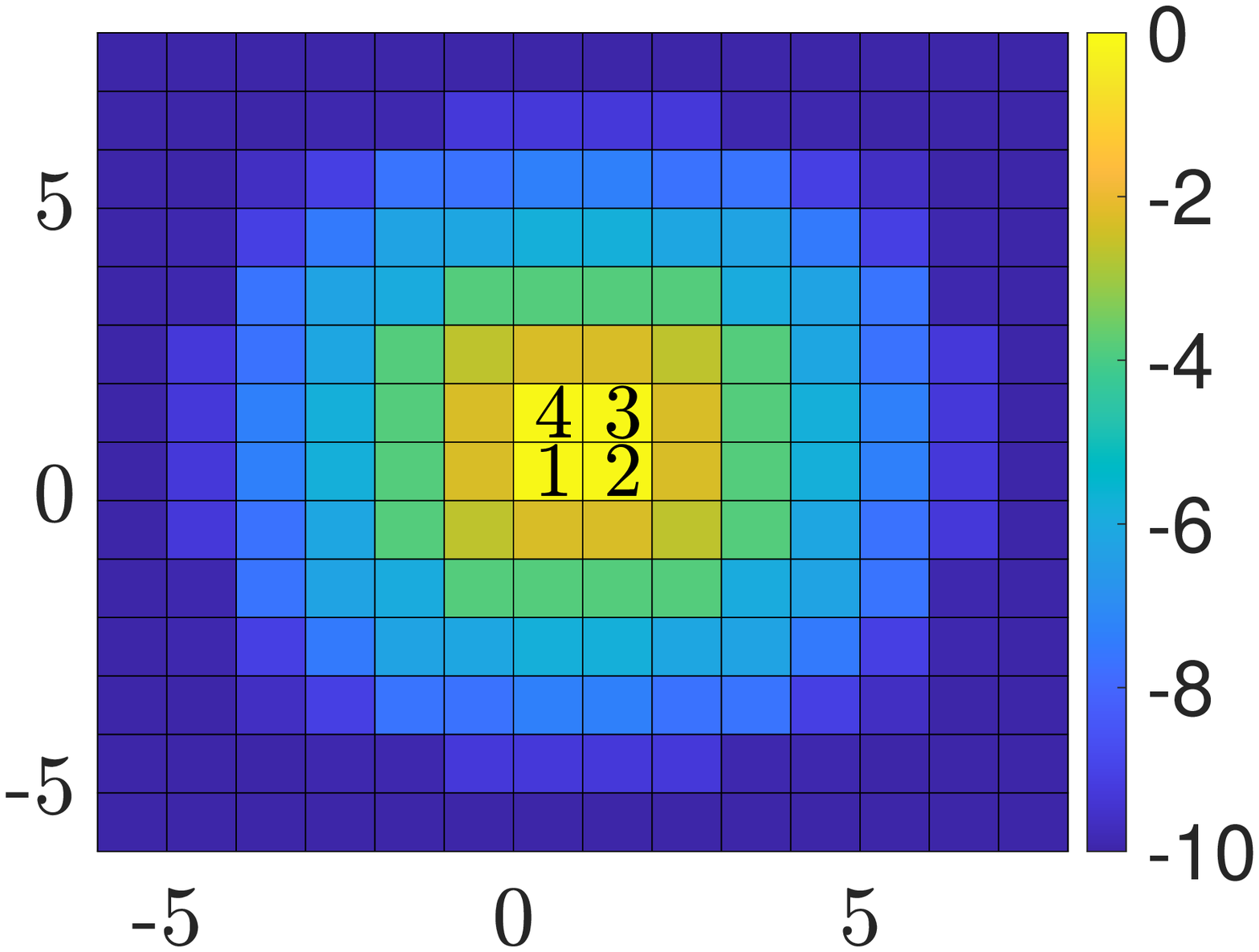}
    \end{subfigure}
    \begin{subfigure}{0.49\linewidth}
    \caption{}
    \label{fig:fundbreatherb}
    \includegraphics[width=4.1cm]{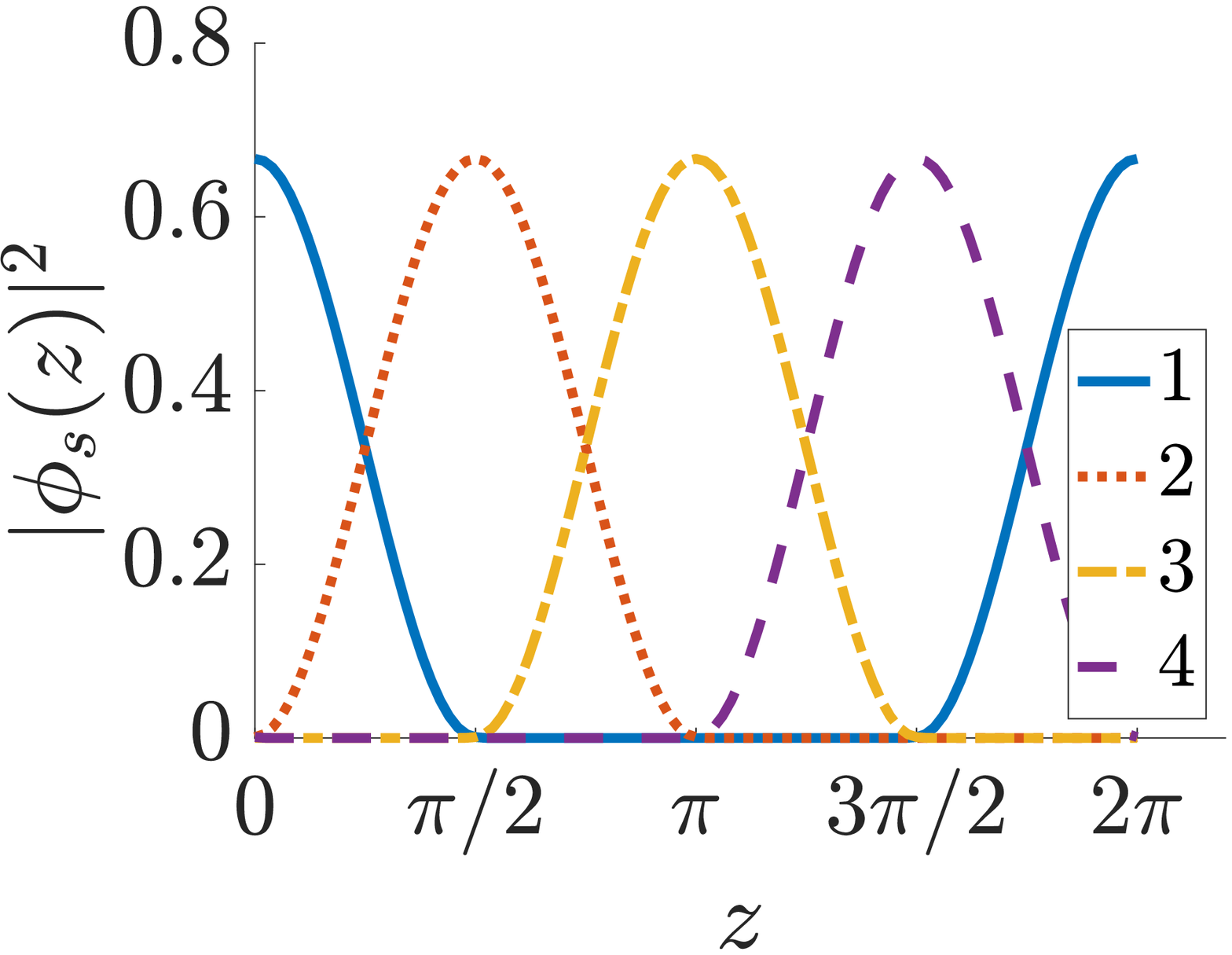}
    \end{subfigure}
    \begin{subfigure}{0.49\linewidth}
    \caption{}
    \label{fig:fundbreatherc}
    \includegraphics[width=4.1cm]{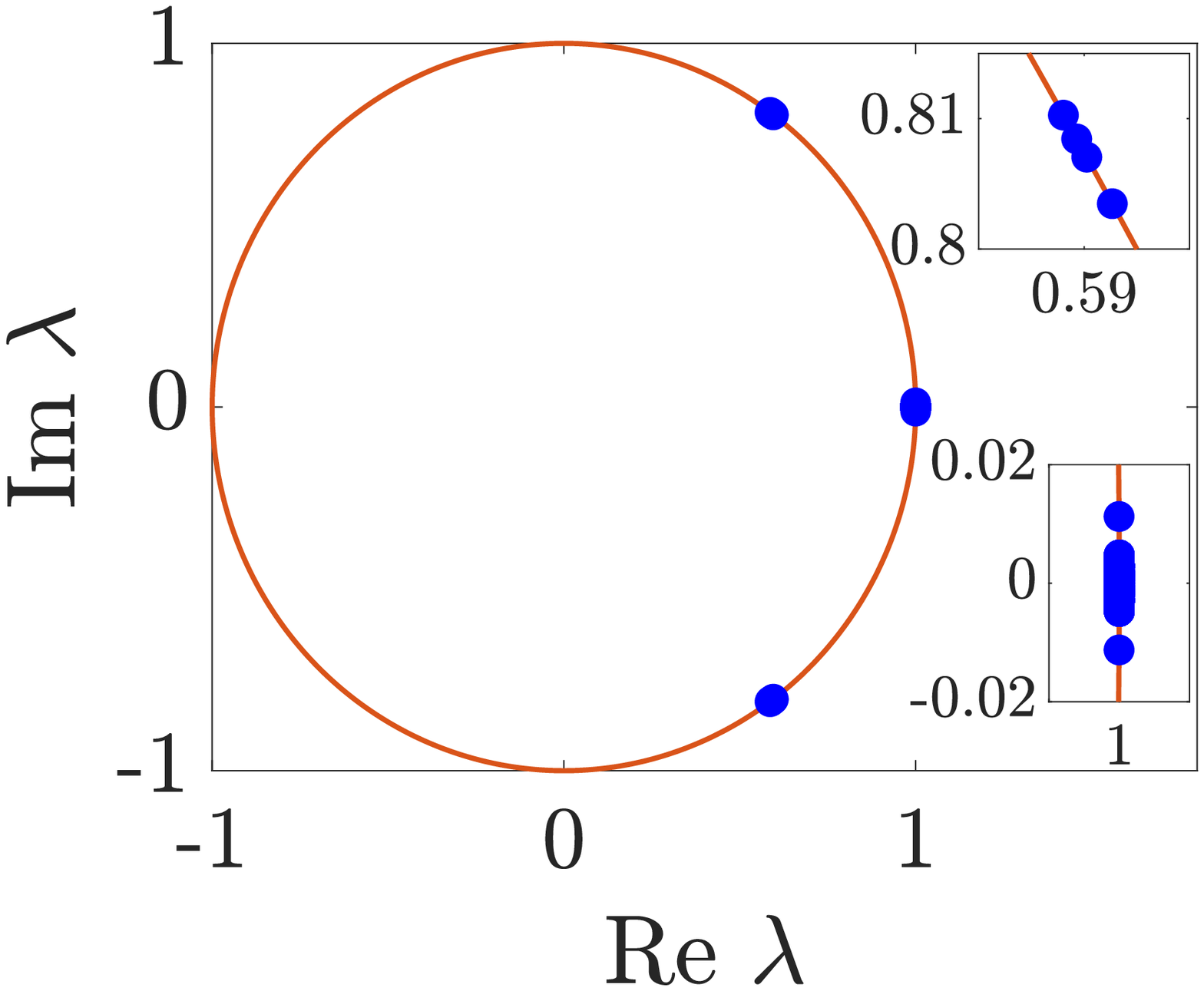}
    \end{subfigure}
    \begin{subfigure}{0.49\linewidth}
    \caption{}
    \label{fig:fundbreatherd}
    \includegraphics[width=4.1cm]{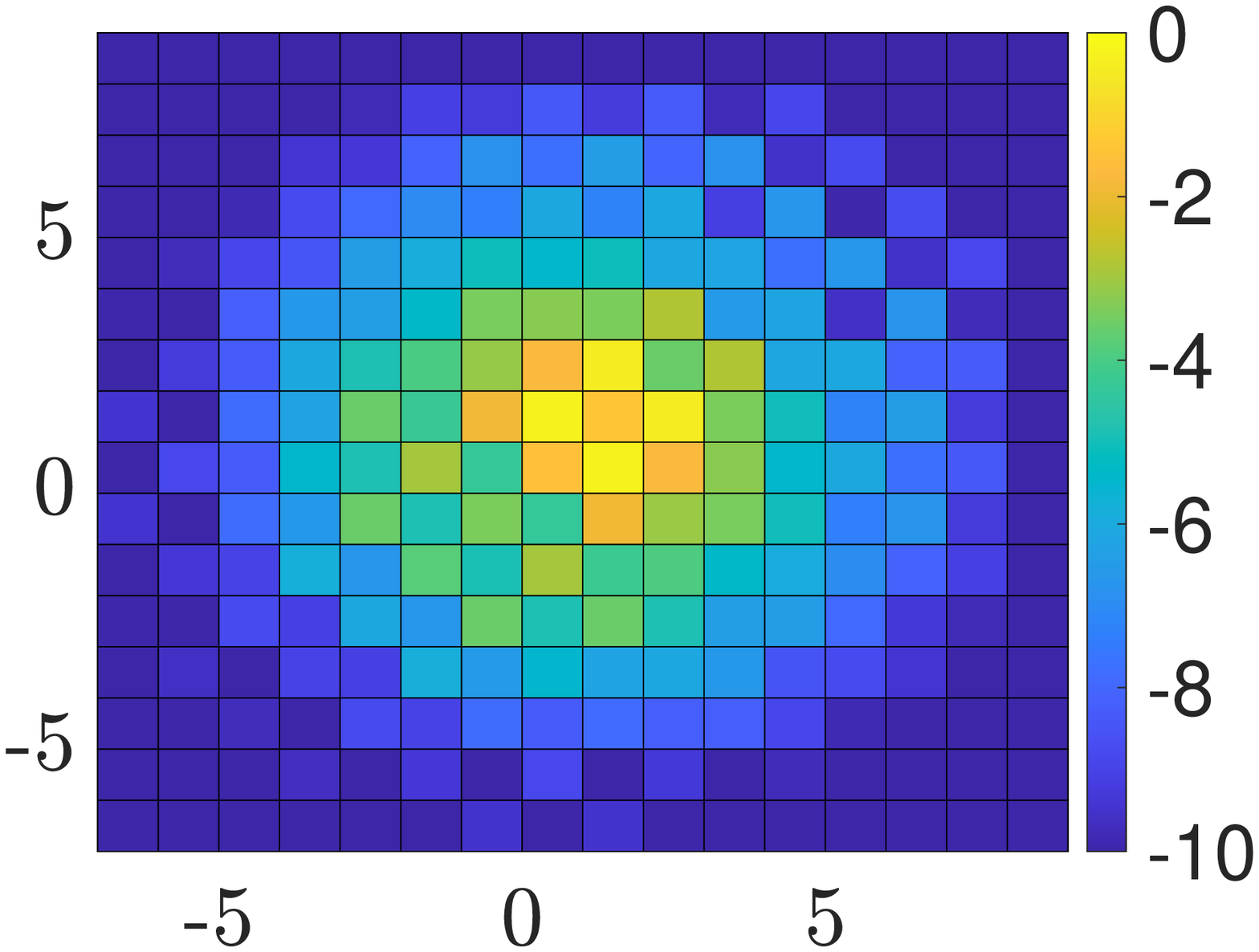}
    \end{subfigure}
    \caption{(a) Color map of the $\log_{10}$ 
    max intensity at each site of the fundamental breather over one period. (b) Intensity of the solution at the four central sites of the fundamental breather over one period $T$; the four sites are labeled in (a). (c) Floquet spectrum of the fundamental breather, with top inset showing the four isolated modes associated with the point spectrum of the solution, and bottom inset detailing the spectrum near (1,0). (d) Color map of the $\log_{10}$ intensity of the isolated Floquet mode at $z=0$ of the fundamental breather solution corresponding to $\lambda = 0.5953 + 0.8035i$; intensity maps of the other isolated Floquet modes are similar. $20\times 20$ lattice, coupling constant $C = 1$, period $T = 2\pi$, phase shift $\theta = \pi$, steepness factor $k = 10$.}
    \label{fig:fundbreather}
\end{figure}

\begin{figure}
    \centering
    \begin{subfigure}{0.49\linewidth}
    \caption{}
    \label{fig:fundspeca}
    \includegraphics[width=4.1cm]{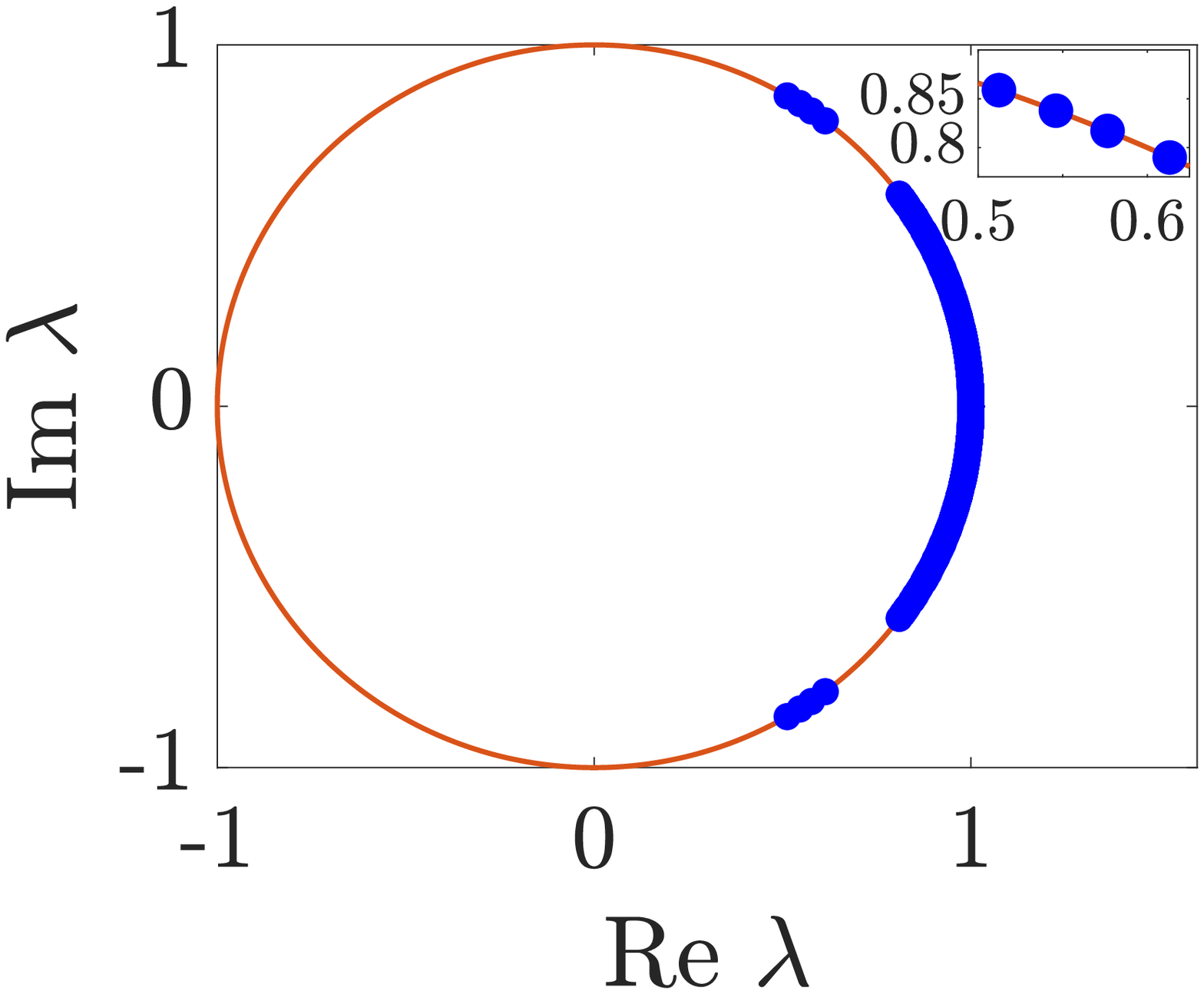}
    \end{subfigure}
    \begin{subfigure}{0.49\linewidth}
    \caption{}
    \label{fig:fundspecb}
    \includegraphics[width=4.1cm]{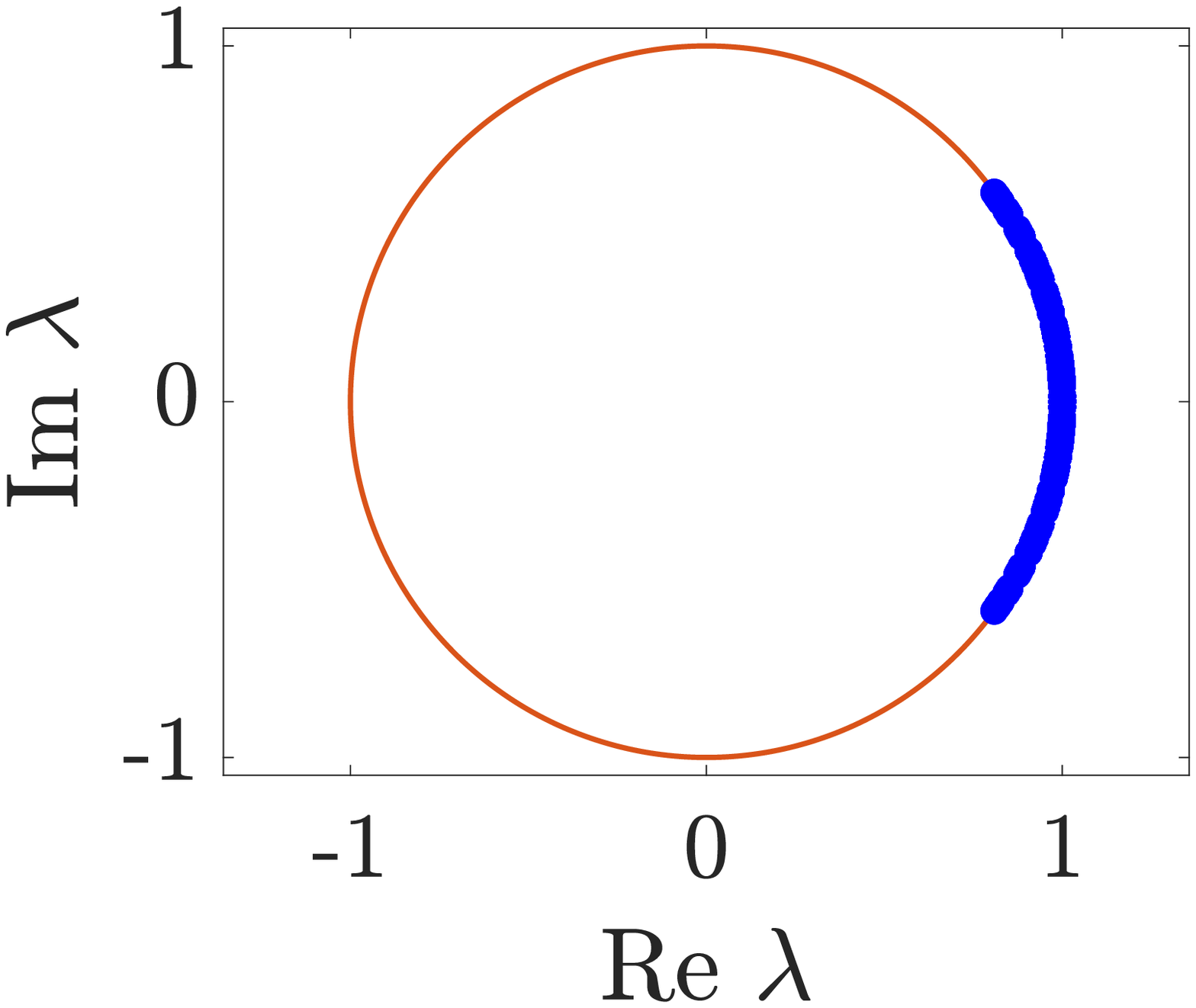}
    \end{subfigure}
    \caption{(a) Floquet spectrum of the fundamental breather solution with inset showing the isolated modes. (b) Floquet spectrum of background state. $20\times 20$ lattice, coupling constant $C=1.05$, period $T = 2\pi$, phase shift $\theta = \pi$, steepness factor $k = 10$.}
    \label{fig:fundspec}
\end{figure}

First, we consider the case when the phase shift is given by $\theta = \pi$. \cref{fig:fundbreathera} and \cref{fig:fundbreatherb} show the fundamental breather solution for coupling strength $C = 1$ and phase shift $\theta = \pi$, which is exponentially localized at the four central sites of the lattice. The Floquet spectrum for $C = 1$ is located on the unit circle (relative error less than $10^{-5}$), which suggests that this solution is stable. There is a small continuous spectrum band near $(1,0)$ on the unit circle (bottom inset in \cref{fig:fundbreatherc}); as the steepness parameter $k$ of the coupling functions $J_j$ increases, the continuous spectrum band for $C=1$ approaches a single point at $(1,0)$. We explore this in \cref{app:bands}, which presents 
a relevant discussion as we approximate $J_j$ by step functions.
In addition, there is a set of four isolated Floquet eigenmodes (top inset in \cref{fig:fundbreatherc}, as well as \cref{fig:fundspeca}), together with their complex conjugates. 
These isolated modes are spatially localized (\cref{fig:fundbreatherd}), as opposed to the nonlocalized continuous spectrum modes. 

\begin{figure}
    \centering
    \begin{subfigure}{0.49\linewidth}
    \caption{}
    \label{fig:fundbganglea}
    \includegraphics[width=4.1cm]{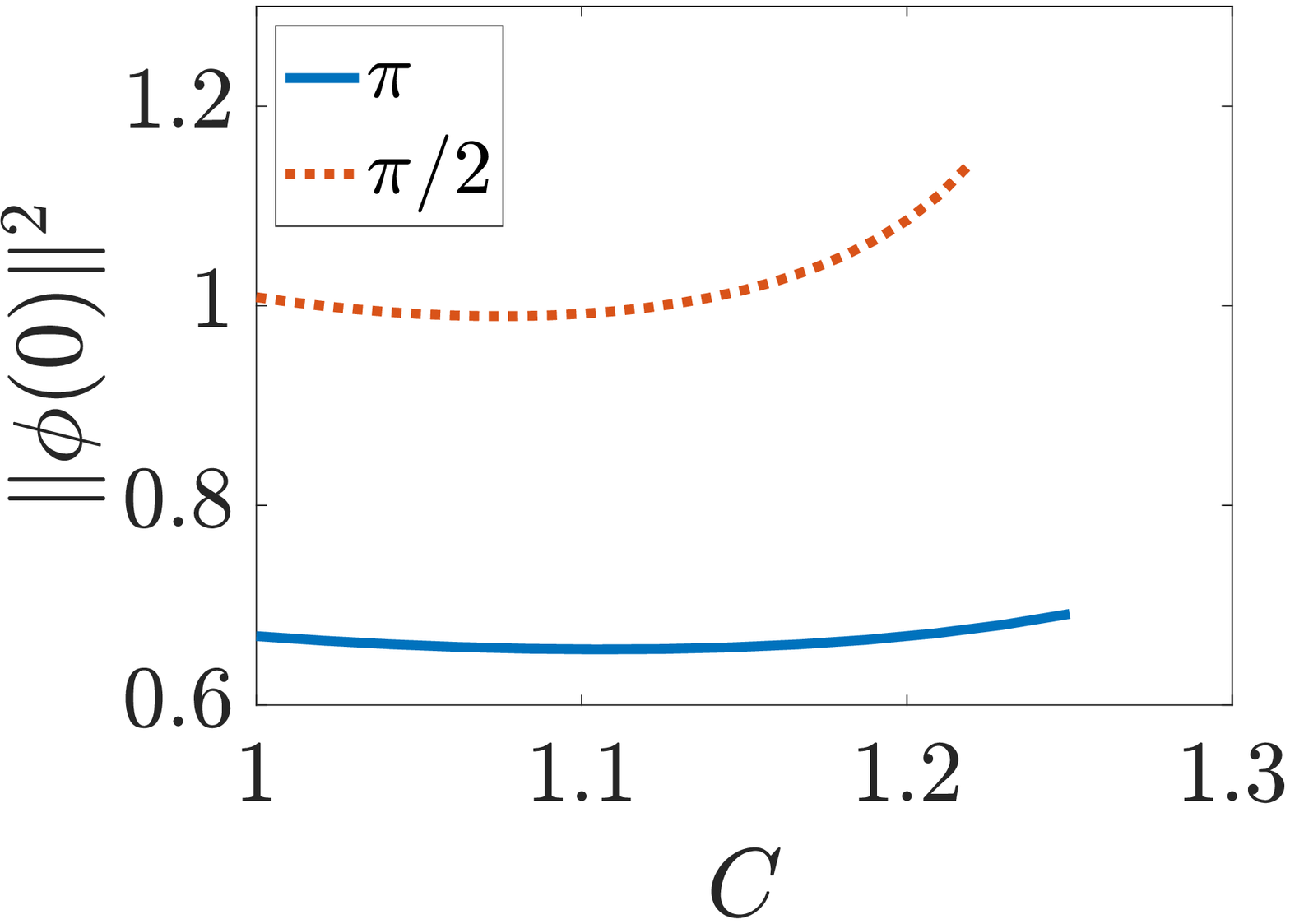}
    \end{subfigure}
    \begin{subfigure}{0.49\linewidth}
    \caption{}
    \label{fig:fundbgangleb}
    \includegraphics[width=4.1cm]{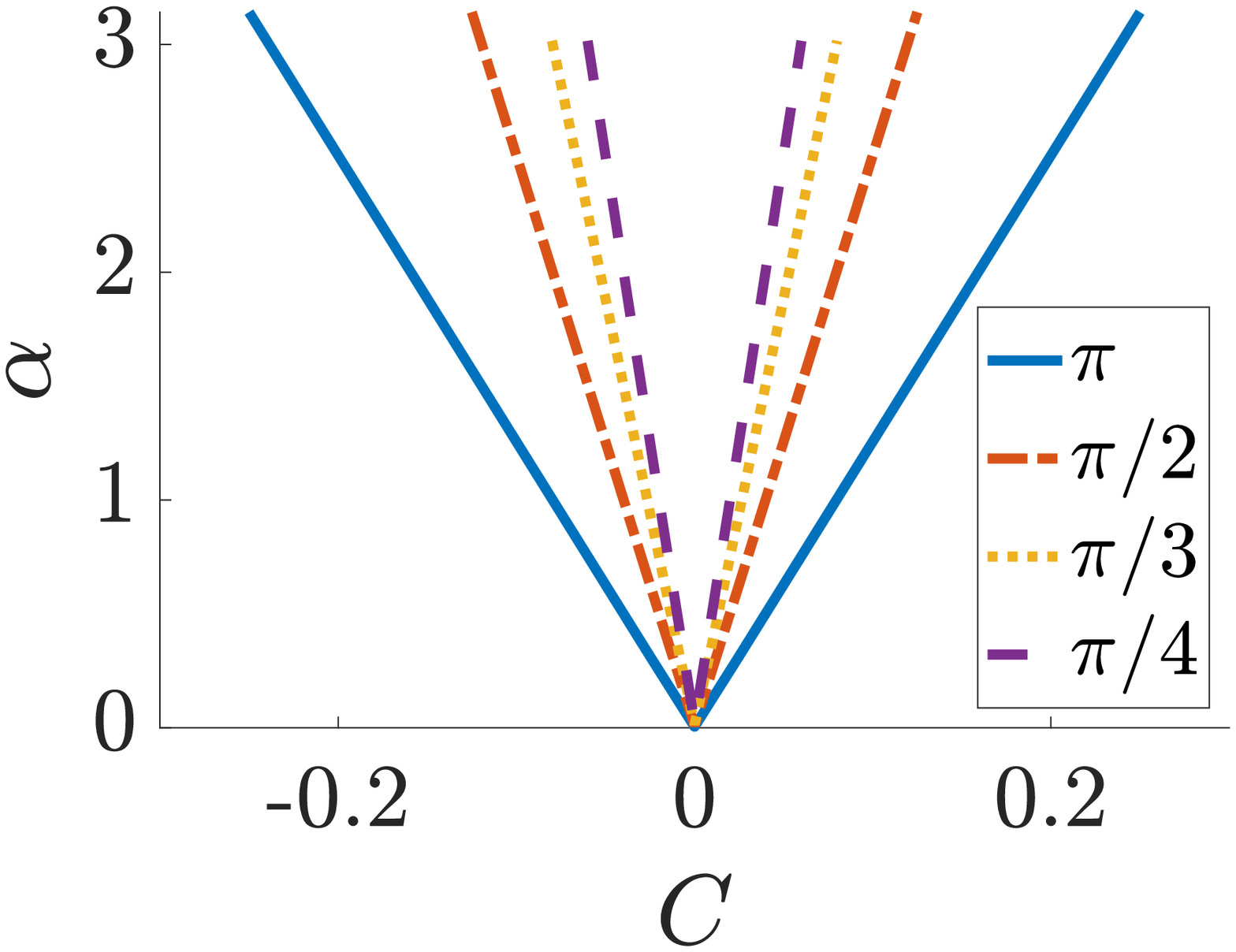}
    \end{subfigure}
    \caption{(a) Norm of solution at $z = 0$ vs. $C$, representing
    the continuation of the fundamental solutions as a function of the
    coupling strength, for phase shifts $\theta =\pi, \pi/2$. (b) Continuous spectrum angle $\alpha$ vs $C$, for phase shifts $\theta =\pi, \pi/2, \pi/3, \pi/4$. This is periodic in $C$, with period $\omega = 2 \pi/T$. $20\times 20$ lattice, coupling period $T = 2\pi$, steepness factor $k = 10$.}
    \label{fig:fundbgangle}
\end{figure}

Let $\sigma_{\text{cont}}(C)$ be the continuous spectrum band, which we compute using the method described in \cref{app:bands}; this agrees with the spectrum determined by computing the eigenvalues of the monodromy matrix. Using this method, we verify numerically that $\sigma_{\text{cont}}(C)$ has the following properties:
\begin{enumerate}[(i)]
\item $\sigma_{\text{cont}}(-C) = \sigma_{\text{cont}}(C)$ (explained by the invariance $u(z;C) \rightarrow u^*(z;-C)$ in equation \cref{eq:linmodel}).
\item $\sigma_{\text{cont}}(C + \omega) = \sigma_{\text{cont}}(C)$, where $\omega = 2 \pi/T$ is the fundamental frequency of the system.
\item For $C \in [-\omega/2, \omega/2]$, $\sigma_c \in [-\alpha, \alpha]$, where $\alpha = C \tau$ (\cref{fig:fundbgangleb}). This extends periodically in $C$ with period $\omega$ for $C$ outside this interval. The band is filled out as the lattice size increases, and is a continuum for the limiting lattice $\Z^2$.
\end{enumerate}
In particular, since the continuous spectrum band is a single point at $(1,0)$ when $C=1$, it follows from the $\omega$-periodicity of the bands that they collapse into a single point at $(1,0)$ (in the limiting case when the coupling functions $J_j$ are step functions) whenever $C$ is an integer multiple of $\omega$. This explains what occurs at $C=1$ in \cref{fig:fundbreatherc}.

\begin{figure}
    \centering
    \begin{subfigure}{0.49\linewidth}
    \caption{}
    \label{fig:fundincreaseCa}
    \includegraphics[width=4.1cm]{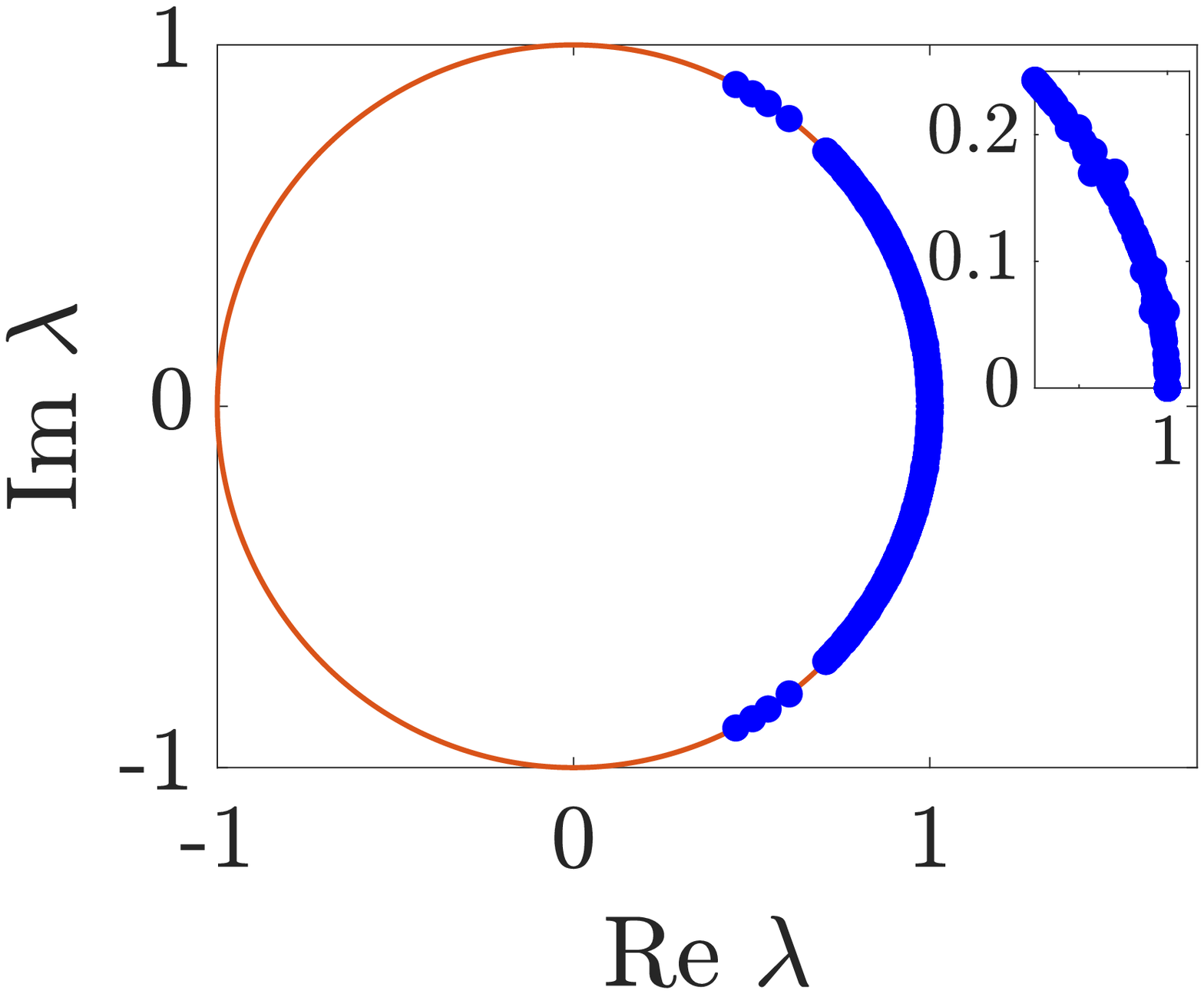}
    \end{subfigure}
    \begin{subfigure}{0.49\linewidth}
    \caption{}
    \label{fig:fundincreaseCb}
    \includegraphics[width=4.1cm]{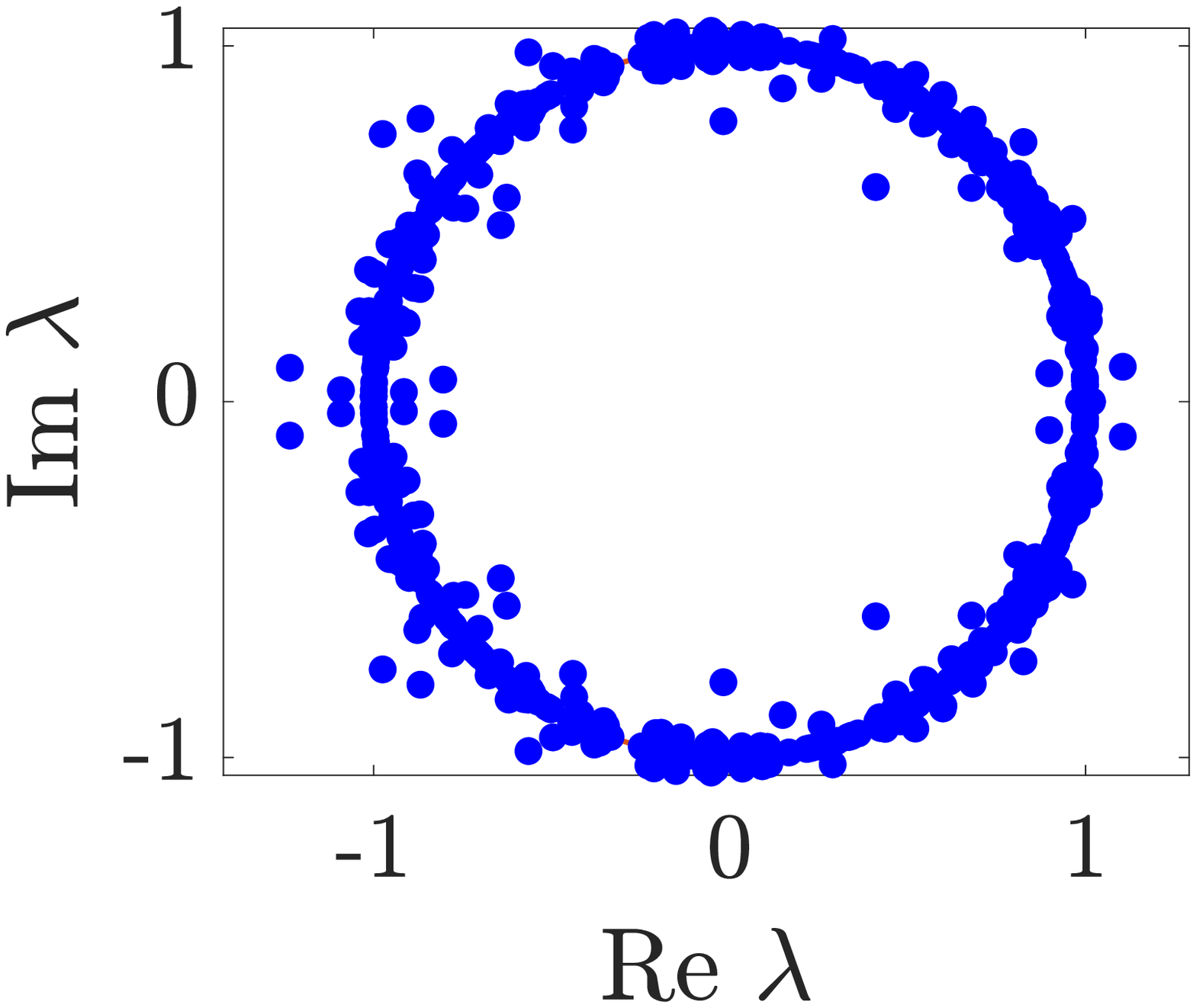}
    \end{subfigure}
    \begin{subfigure}{0.49\linewidth}
    \caption{}
    \label{fig:fundincreaseCc}
    \includegraphics[width=4.1cm]{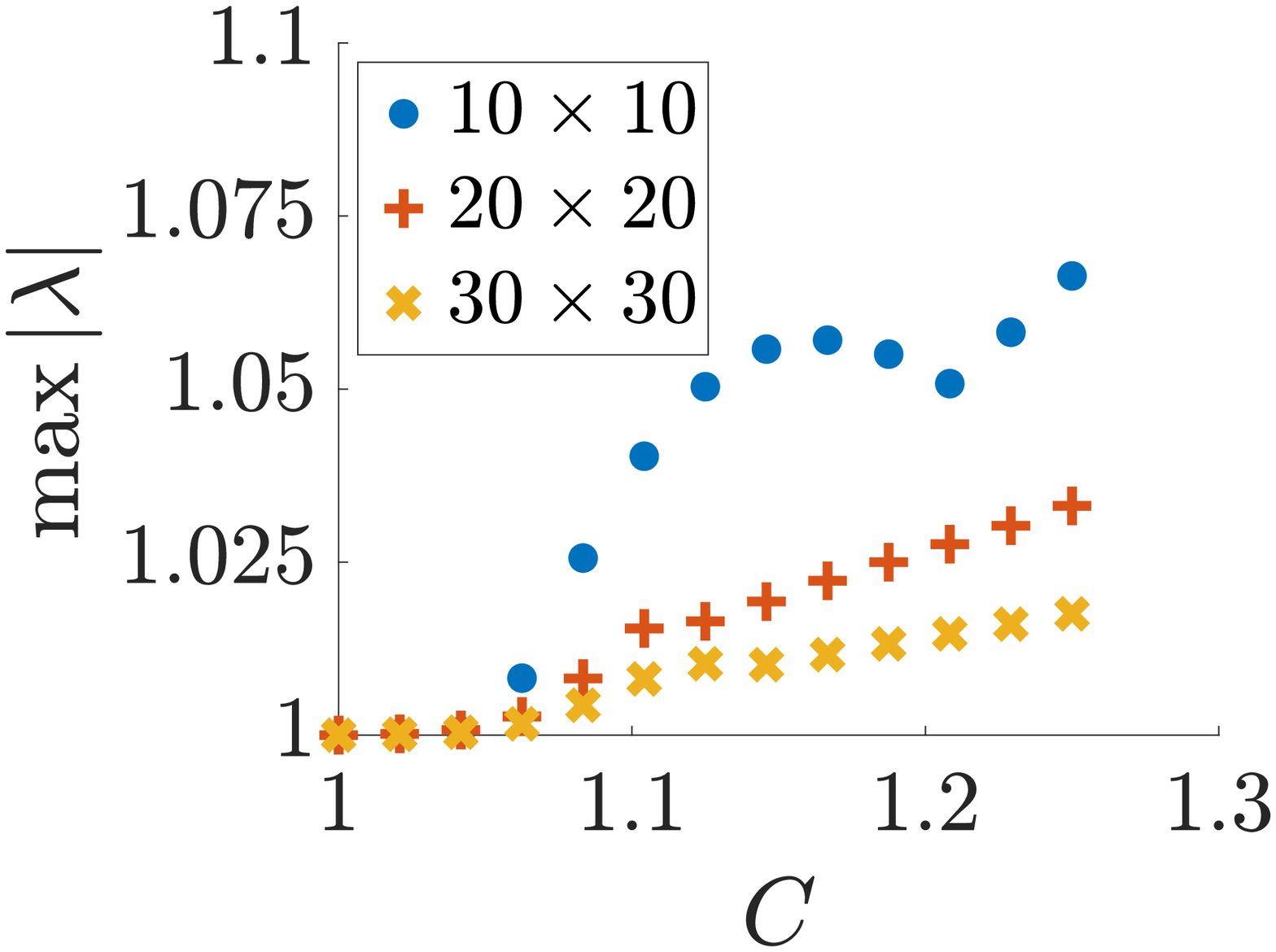}
    \end{subfigure}
    \begin{subfigure}{0.49\linewidth}
    \caption{}
    \label{fig:fundincreaseCd}
    \includegraphics[width=4.1cm]{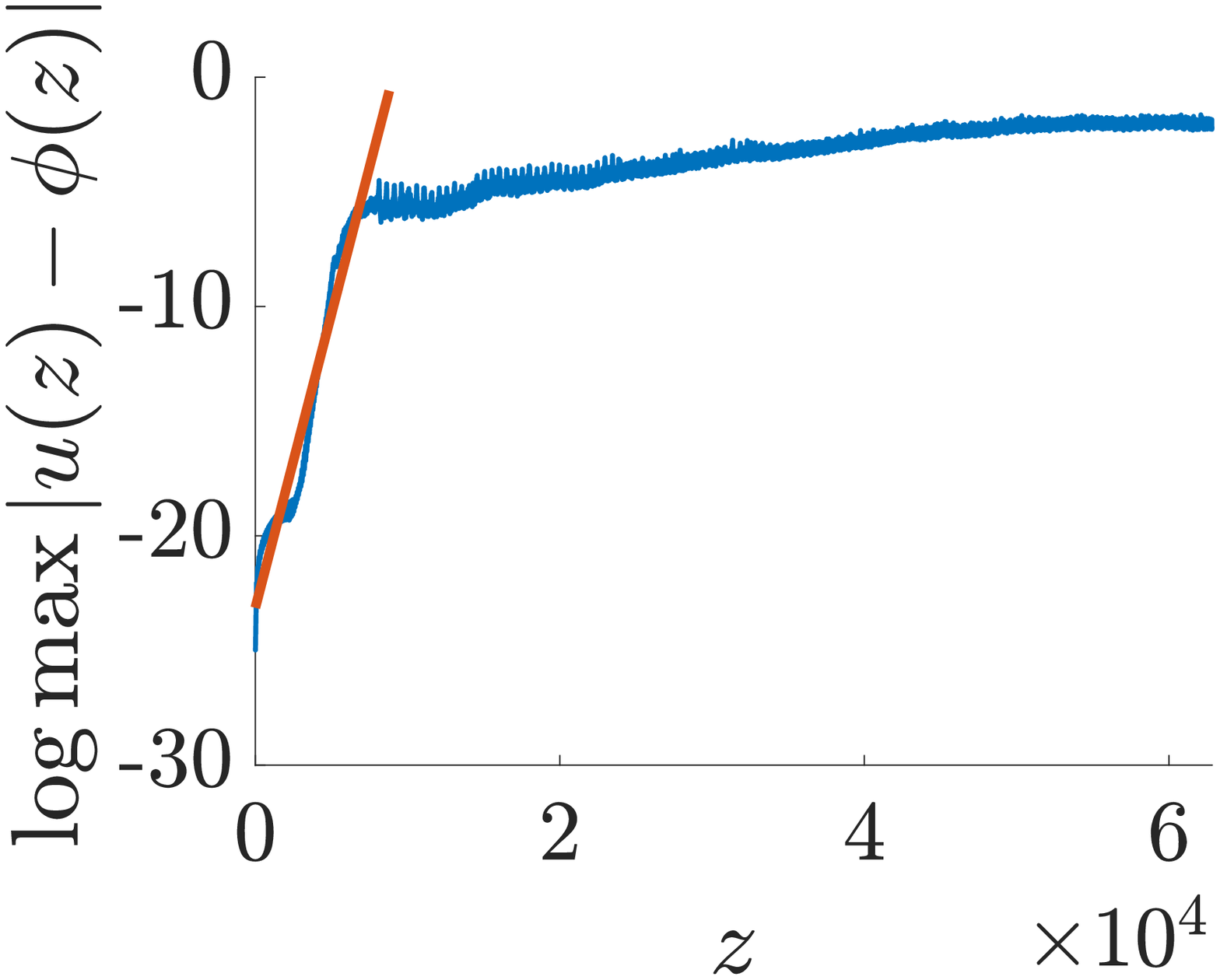}
    \end{subfigure}
    \caption{(a)-(b) Floquet spectrum of the fundamental breather for $C=1.0625$ (a) and $C=1.5$ (b). (c) Maximum absolute value of Floquet multiplier $\lambda$ vs $C$ for increasing grid size. (d) Log of the maximum absolute difference between the perturbed solution $u(z)$ and the fundamental breather $\phi(z)$; initial condition for perturbation is $\phi(0)+\epsilon v$, where $\epsilon = 10^{-10}$ and $v$ is the eigenfunction corresponding to the largest Floquet multiplier ($|\lambda| = 1.033$ for $C = 1.25$). The line represents the least squares regression line of the early growth stage of the dynamics. 
    $20 \times 20$ lattice, period $T = 2\pi$, phase shift $\theta = \pi$, steepness factor $k = 10$.}
    \label{fig:fundincreaseC}
\end{figure}

Using parameter continuation, we can compute solutions for other values of $C$ by gradually increasing (or decreasing) $C$ (\cref{fig:fundbganglea}). 
It is relevant to note here that
in the experimental setting of interest~\cite{Mukherjee2020}, by controlling the separation
    between waveguides during the fabrication process, it is,
    in principle, possible to control the lattice spacing,
    and, accordingly, to enable the manifestation of the
    dynamical features reported herein.
As $C$ is increased from 1, some Floquet multipliers in the continuous spectrum band collide and leave the unit circle (\cref{fig:fundincreaseCa} and \cref{fig:fundincreaseCb}). In particular, we see from \cref{fig:fundincreaseCa} that the isolated modes are not involved in these collisions. Although the Floquet multiplier with maximum absolute value increases with $C$, the rate of growth decreases as the lattice size increases (\cref{fig:fundincreaseCc}); the latter
suggests that this type of instability disappears in the infinite lattice limit, in line, e.g., with what is 
known from the classic work of~\cite{Marin1998}. 
We can see the consequences of this unstable Floquet eigenmode by perturbing the fundamental breather  with a small multiple of the eigenfunction corresponding to the largest Floquet multiplier (\cref{fig:fundincreaseCd}). The slope of the least squares regression line in the figure is within 2\% of  $\log |\lambda|^{1/\tau}$, where $\lambda$ is the largest Floquet multiplier, confirming the 
results of our spectral stability analysis. At longer times, the solution continues to slowly deviate from the unstable initial condition, yet does not settle into a clearly 
discernible pattern for the $z$ interval of our numerical computations.

\begin{figure}
    \centering
    \begin{subfigure}{0.49\linewidth}
    \caption{}
    \label{fig:fundpertvaryCa}
    \includegraphics[width=4.1cm]{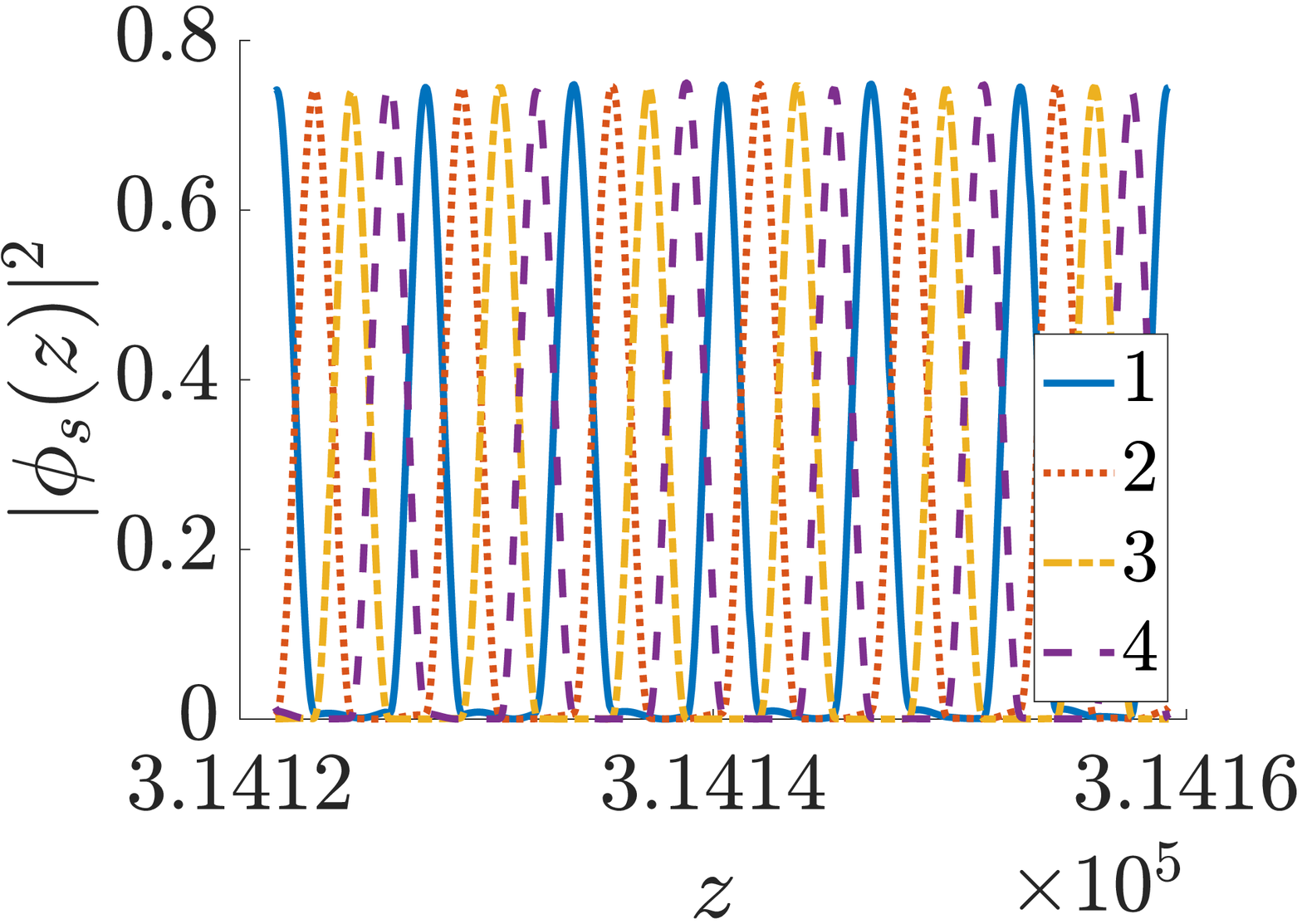}
    \end{subfigure}
    \begin{subfigure}{0.49\linewidth}
    \caption{}
    \label{fig:fundpertvaryCb}
    \includegraphics[width=4.1cm]{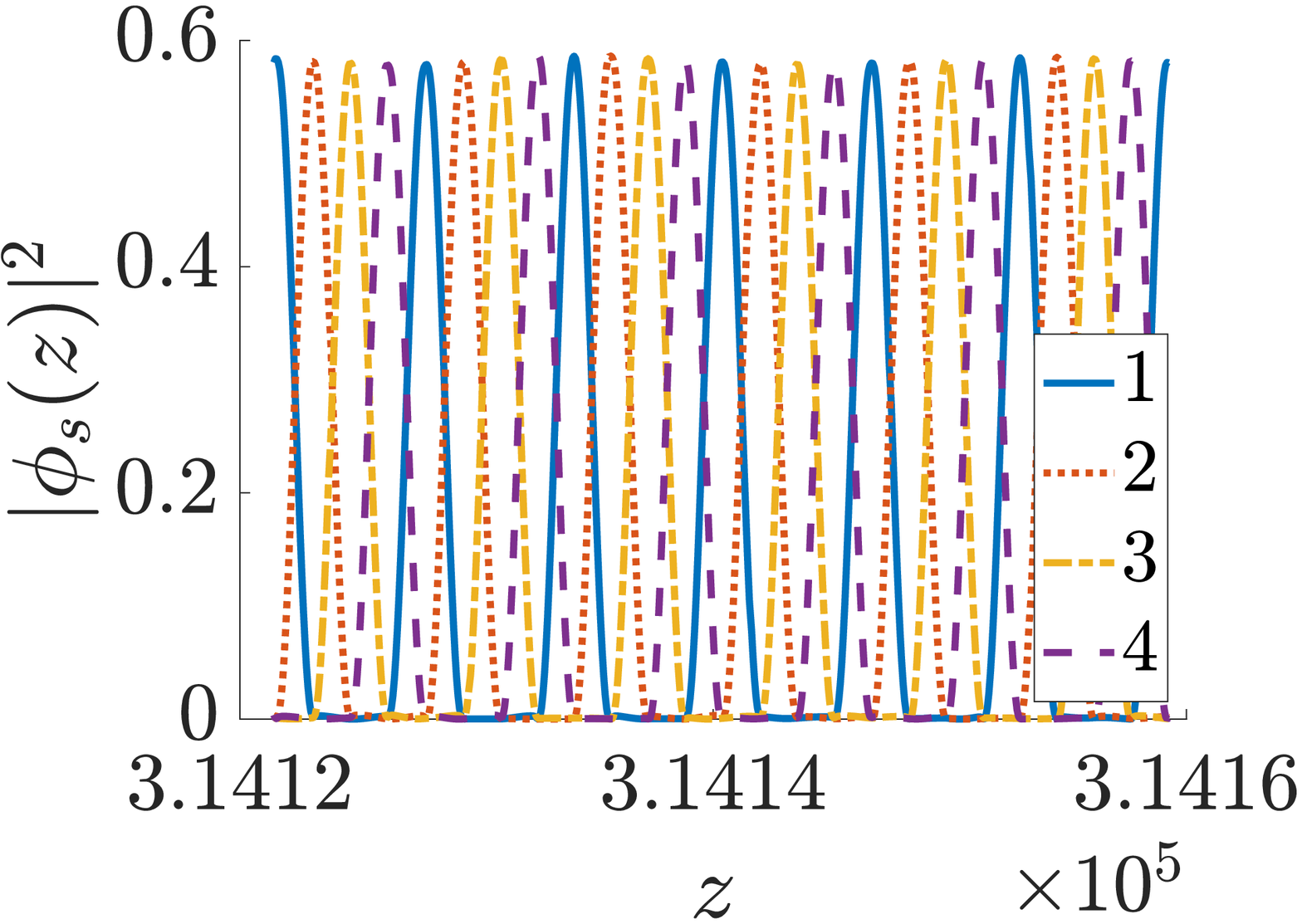}
    \end{subfigure}
    \begin{subfigure}{0.49\linewidth}
    \caption{}
    \label{fig:fundpertvaryCc}
    \includegraphics[width=4.1cm]{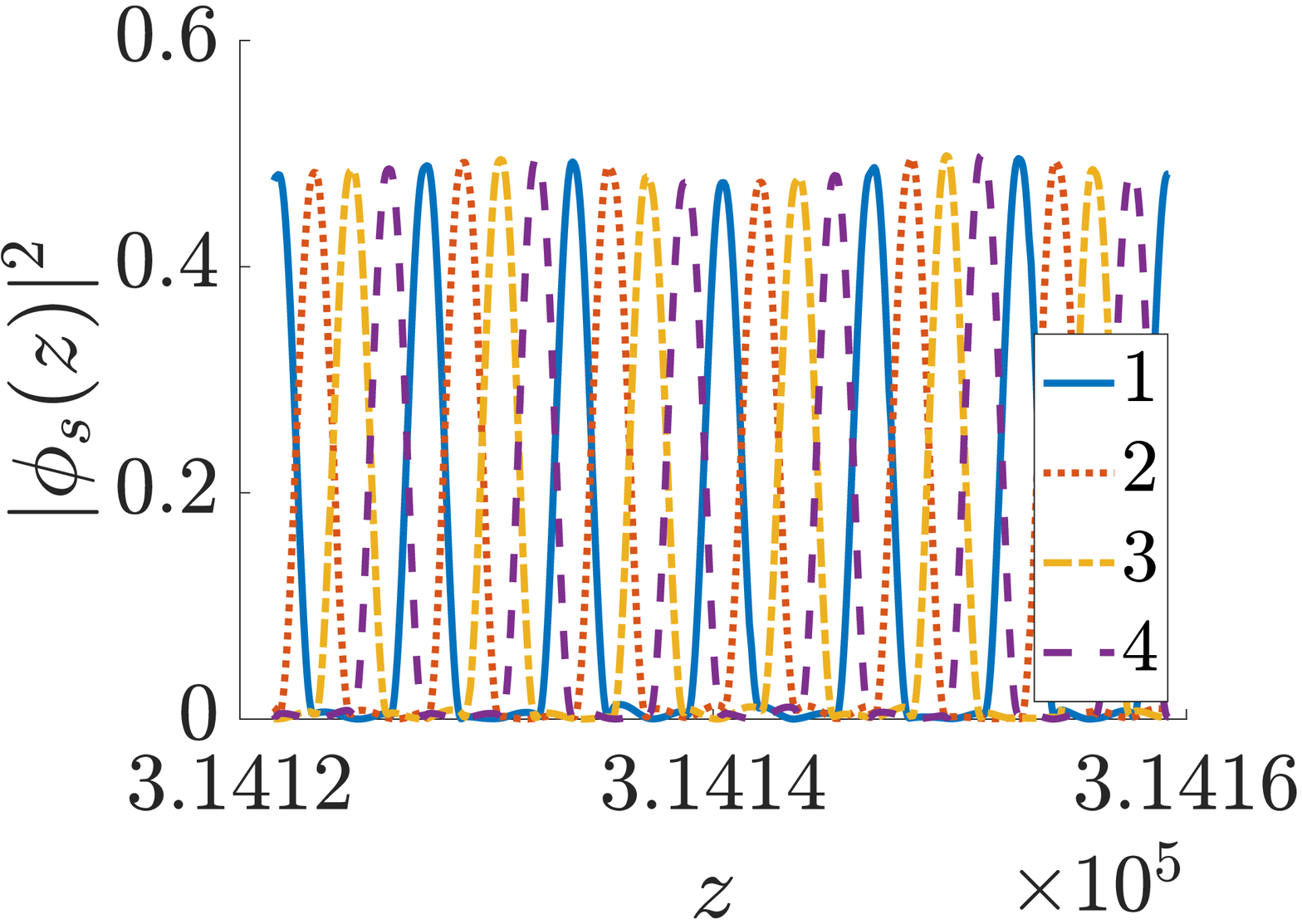}
    \end{subfigure}
    \begin{subfigure}{0.49\linewidth}
    \caption{}
    \label{fig:fundpertvaryCd}
    \includegraphics[width=4.1cm]{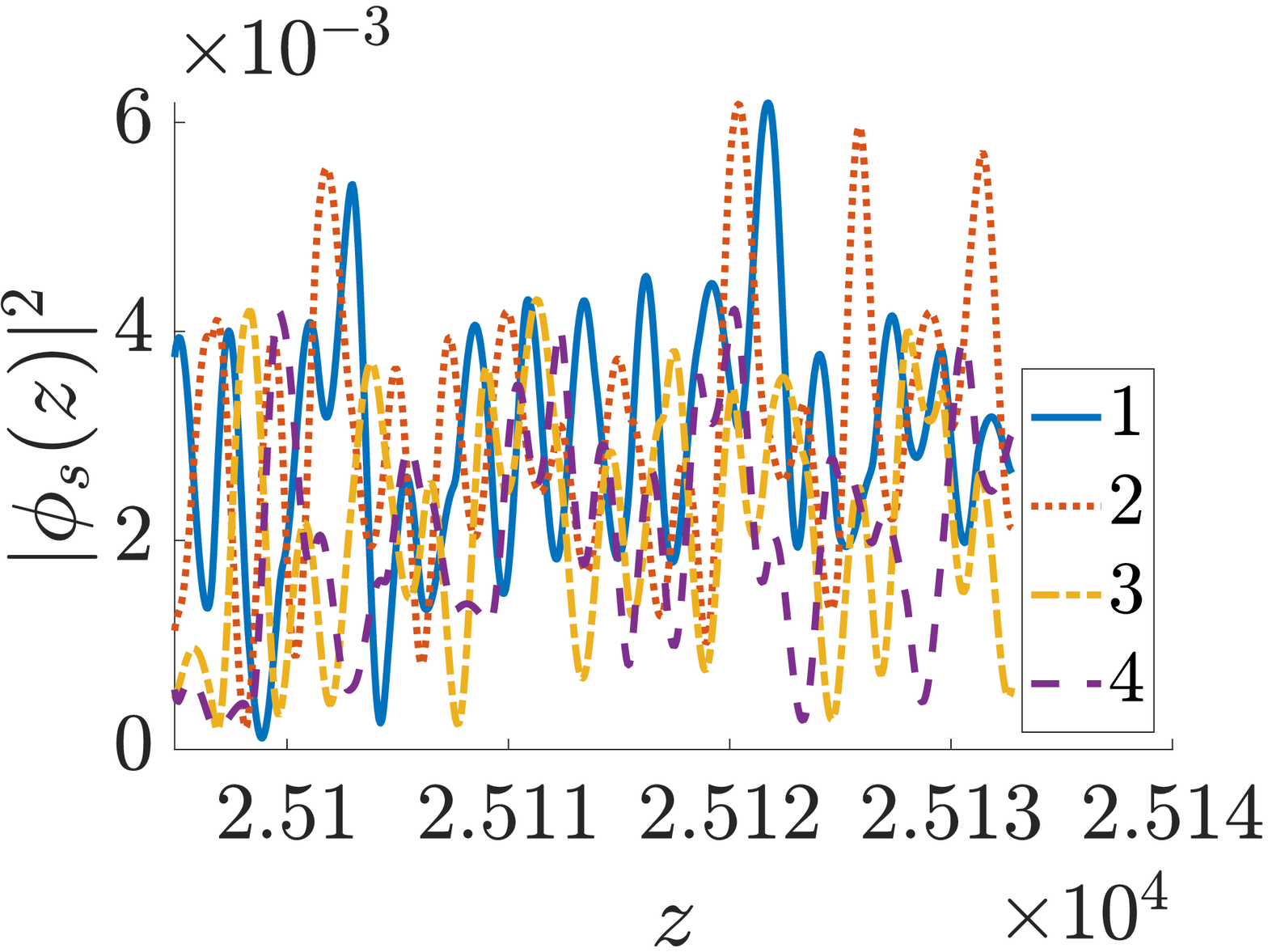}
    \end{subfigure}
    \caption{(a)-(d) Solution at the four central sites for evolution in $z$ of the perturbed fundamental breather $C=1$ (a), $C=1.0625$ (b), $C=1.125$ (c), and $C=1.2083$ (d). Initial condition is a single excited lattice site with the same intensity as the fundamental breather. Fourth order Runge-Kutta scheme, step size $\pi/50$. Lattice size $20\times 20$, period $T = 2\pi$, phase shift $\theta = \pi$, steepness factor $k = 10$.}
    \label{fig:fundpertvaryC}
\end{figure}

Long-term evolution numerical experiments provide further evidence that the fundamental breather solution is stable for $\theta = \pi$ and $C$ close to 1. As an initial condition, we start with all the intensity confined to a single lattice site. The magnitude of this intensity is chosen to be the maximal intensity of the fundamental breather. Results of this evolution for $C=1$, $C=1.0625$, and $C=1.125$ are shown in \cref{fig:fundpertvaryC}. For larger values of $C$, this
initial condition is found to disperse over longer intervals of evolution in $z$ (\cref{fig:fundpertvaryCd}). This effect persists with larger lattice sizes.

\subsection{Phase shift \texorpdfstring{$\theta = \pi/2$}{pi/2}}

\begin{figure}
    \centering
    \begin{subfigure}{0.49\linewidth}
    \caption{}
    \label{fig:fundpi2a}
    \includegraphics[width=4.1cm]{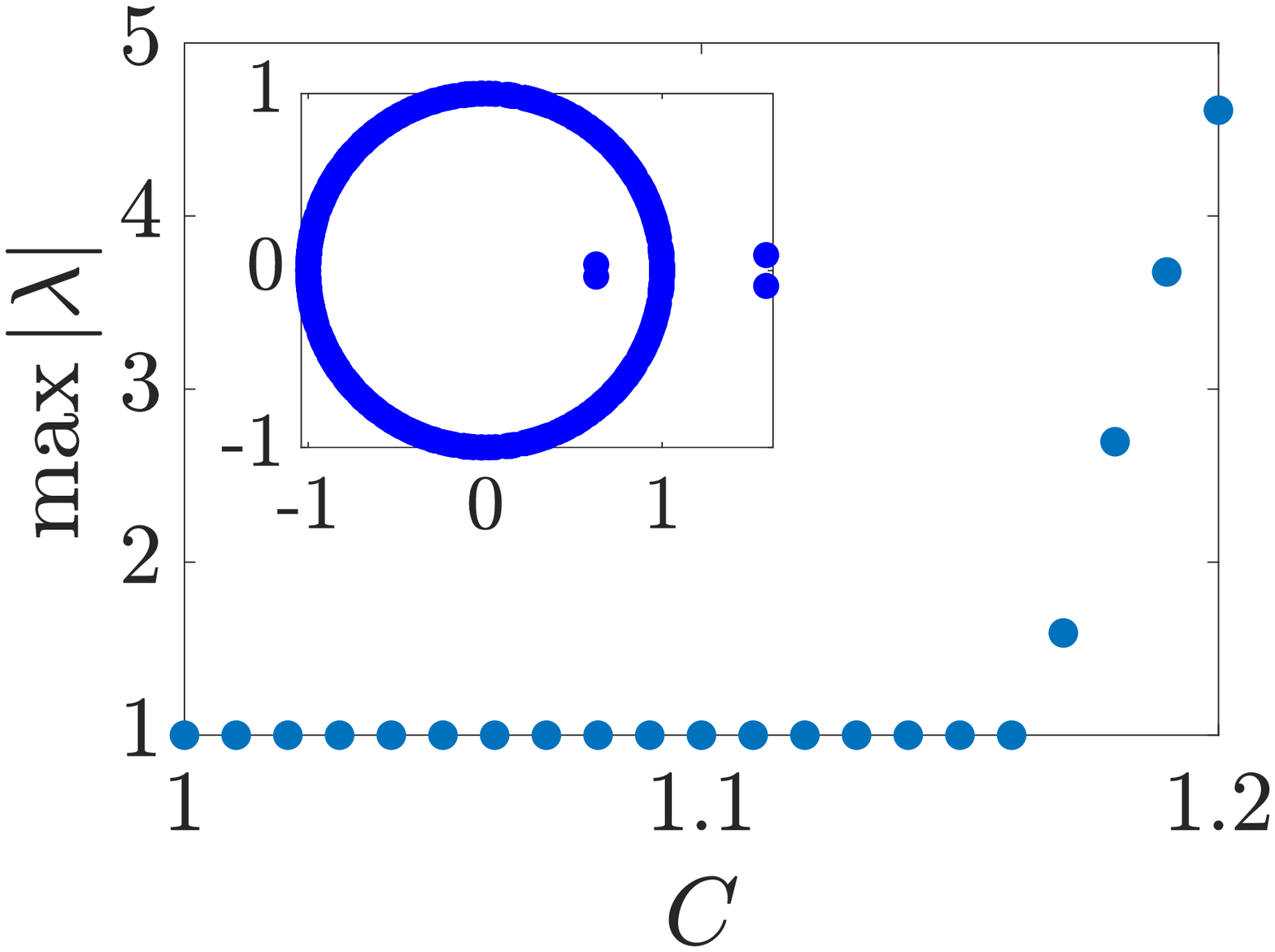}
    \end{subfigure}
    \begin{subfigure}{0.49\linewidth}
    \caption{}
    \label{fig:fundpi2b}
    \includegraphics[width=4.1cm]{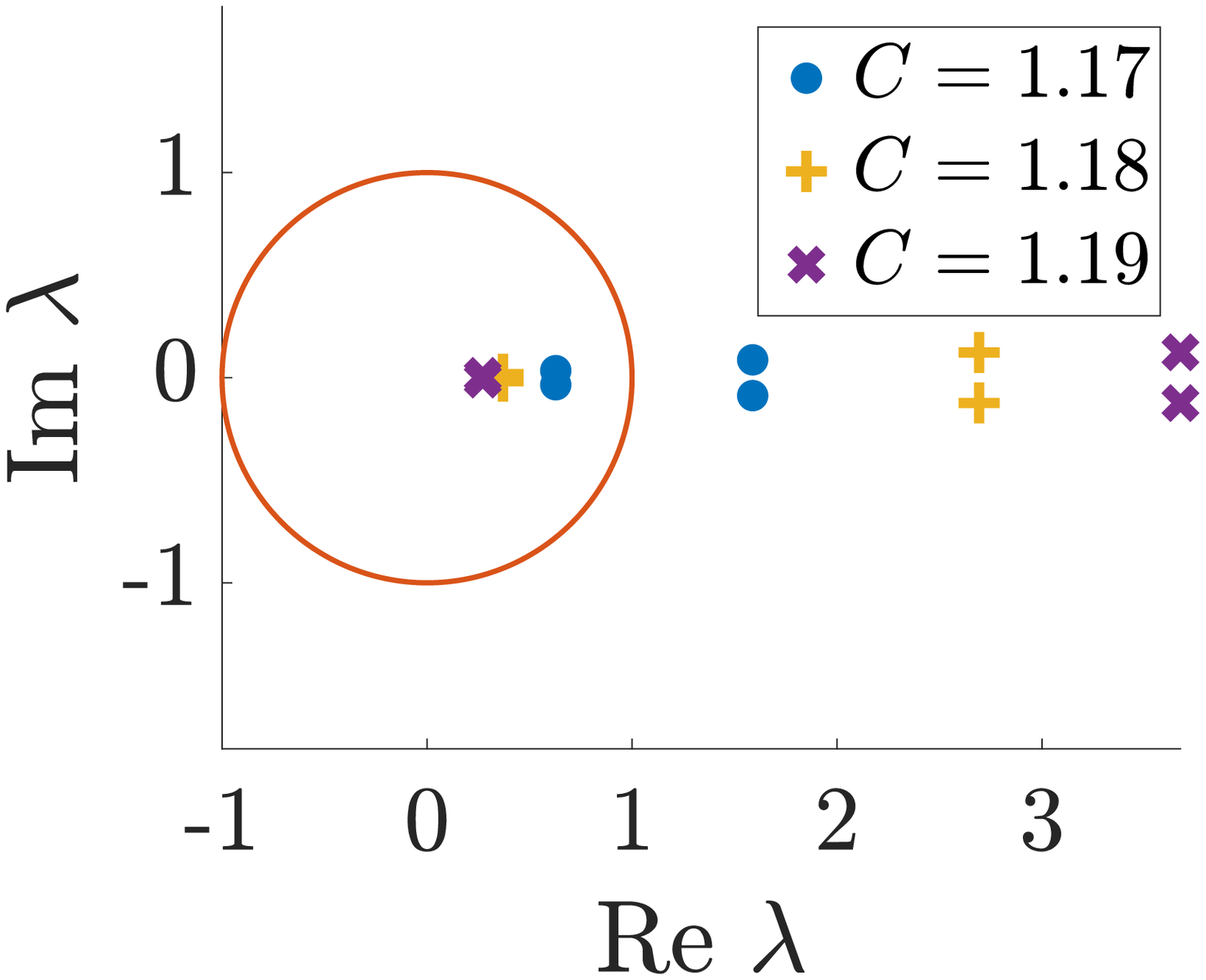}
    \end{subfigure}
    \caption{ (a) Maximum absolute value of Floquet multiplier $\lambda$ vs $C$. Inset shows Floquet spectrum of the fundamental breather for $C=1.17$. (b) Unstable Floquet eigenmodes for $C = 1.17, 1.18$ and $1.19$. $20\times 20$ lattice, period $T = 2\pi$, phase shift $\theta = \pi/2$, steepness factor $k = 10$. }
    \label{fig:fundpi2}
\end{figure}

\begin{figure}
    \centering
    \begin{subfigure}{0.49\linewidth}
    \caption{}
    \label{fig:fundpi2perta}
    \includegraphics[width=4.1cm]{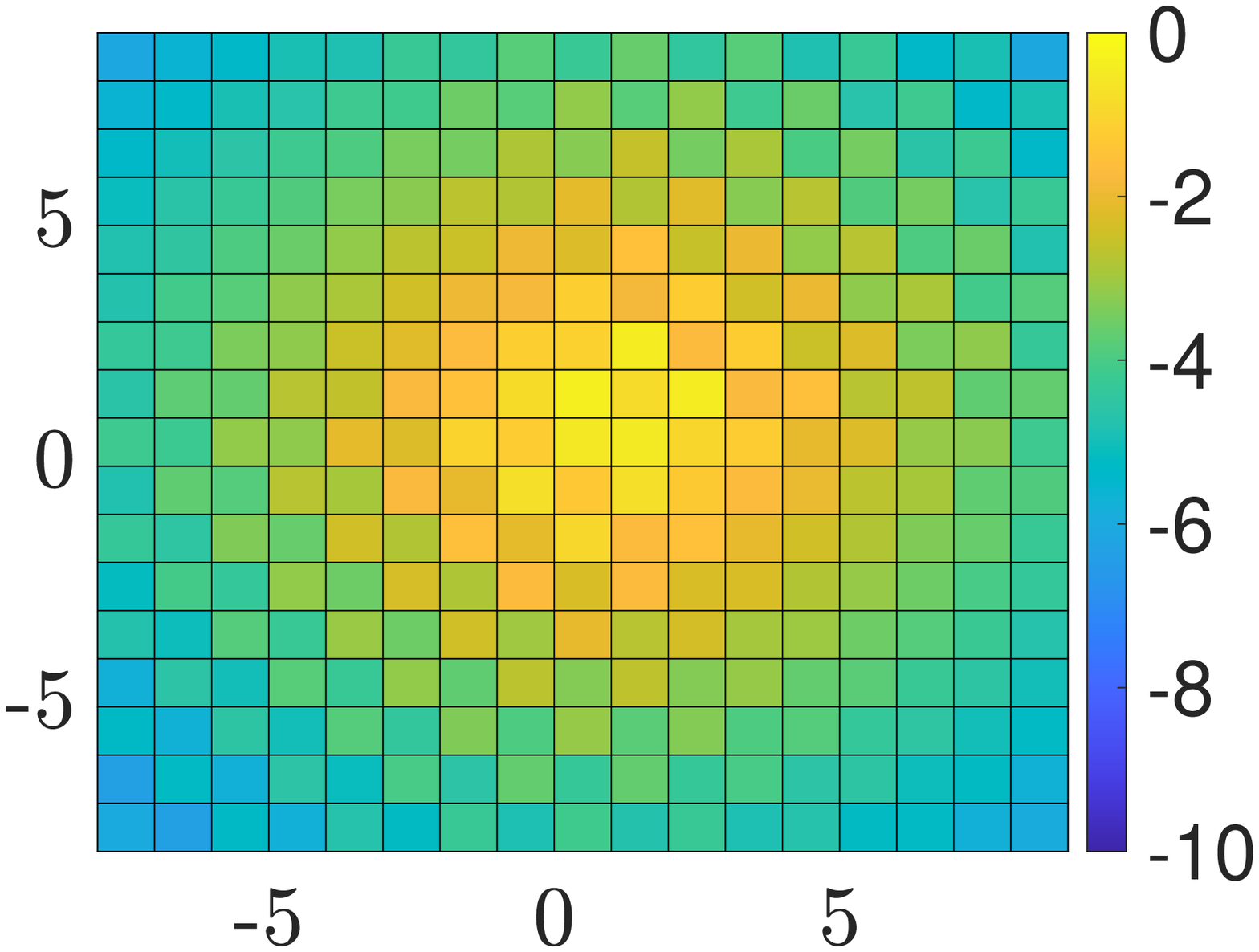}
    \end{subfigure}
    \begin{subfigure}{0.49\linewidth}
    \caption{}
    \label{fig:fundpi2pertb}
    \includegraphics[width=4.1cm]{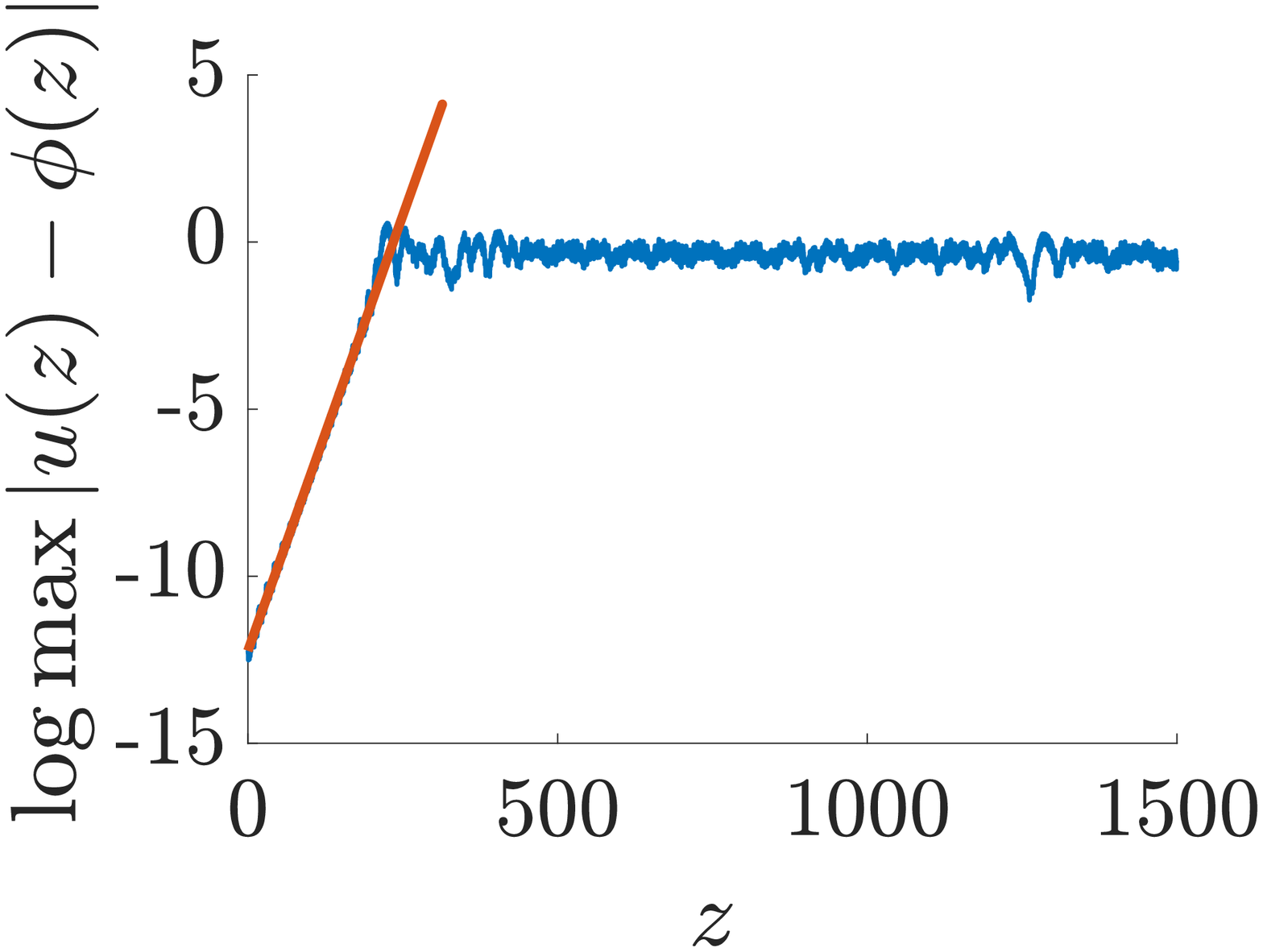}
    \end{subfigure}
    \caption{(a) Color map of the $\log_{10}$ intensity of the unstable Floquet mode ($\lambda = 3.6748+0.1238i$, $|\lambda = 3.6769|$). (b) Log of the maximum absolute difference between the perturbed solution $u(z)$ and the fundamental breather $\phi(z)$; initial condition for perturbation is $\phi(0)+\epsilon v$, where $\epsilon = 10^{-5}$ and $v$ is the eigenfunction corresponding to the most unstable Floquet eigenmode. The line represents the least squares regression fit of the growth portion of the curve. 
    $20 \times 20$ lattice, $C = 1.19$, period $T = 2\pi$, phase shift $\theta = \pi/2$, steepness factor $k = 10$.}
    \label{fig:fundpi2pert}
\end{figure}

\begin{figure}
    \centering
    \begin{subfigure}{0.49\linewidth}
    \caption{}
    \label{fig:fundpi2powera}
    \includegraphics[width=4.1cm]{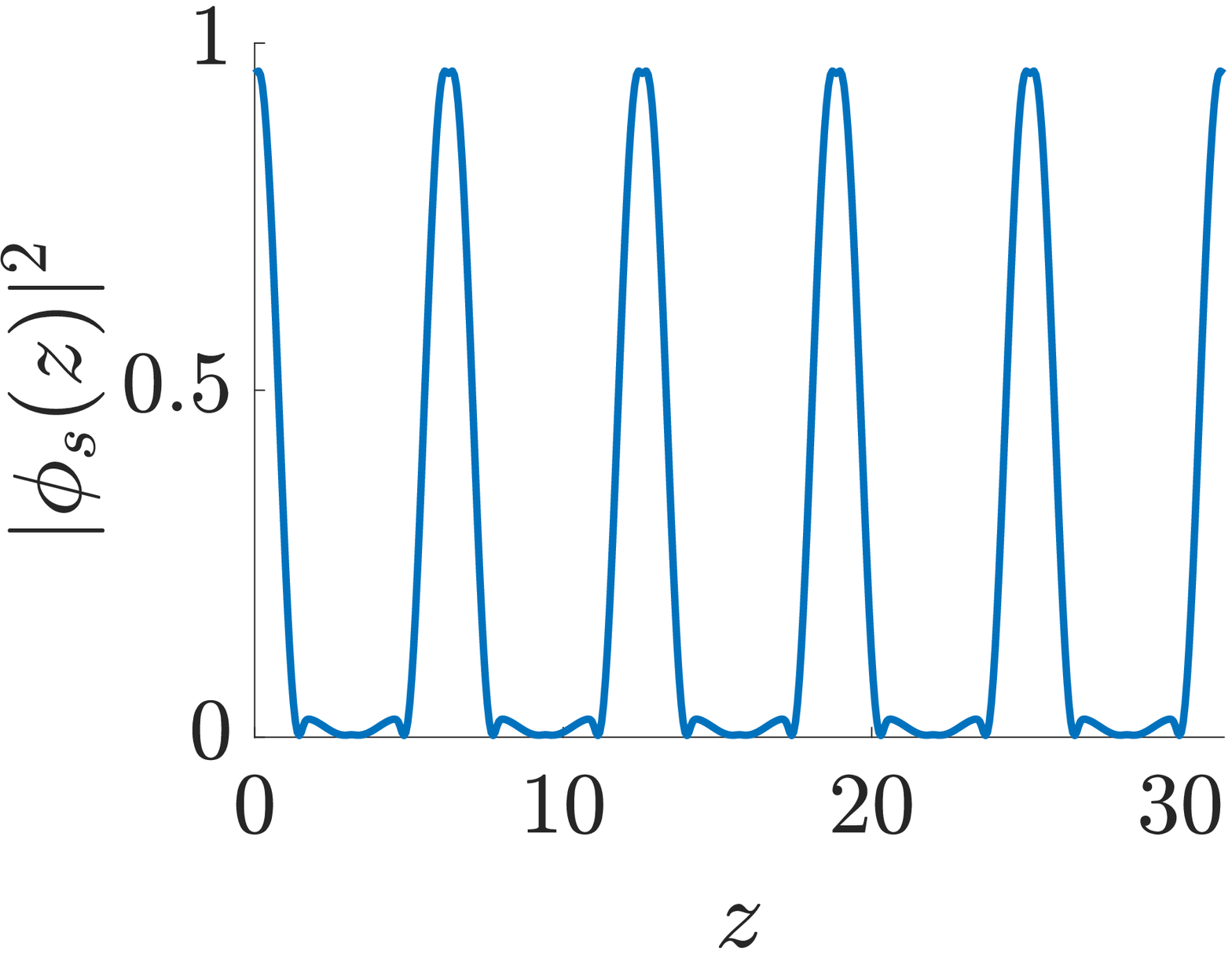}
    \end{subfigure}
    \begin{subfigure}{0.49\linewidth}
    \caption{}
    \label{fig:fundpi2powerb}
    \includegraphics[width=4.1cm]{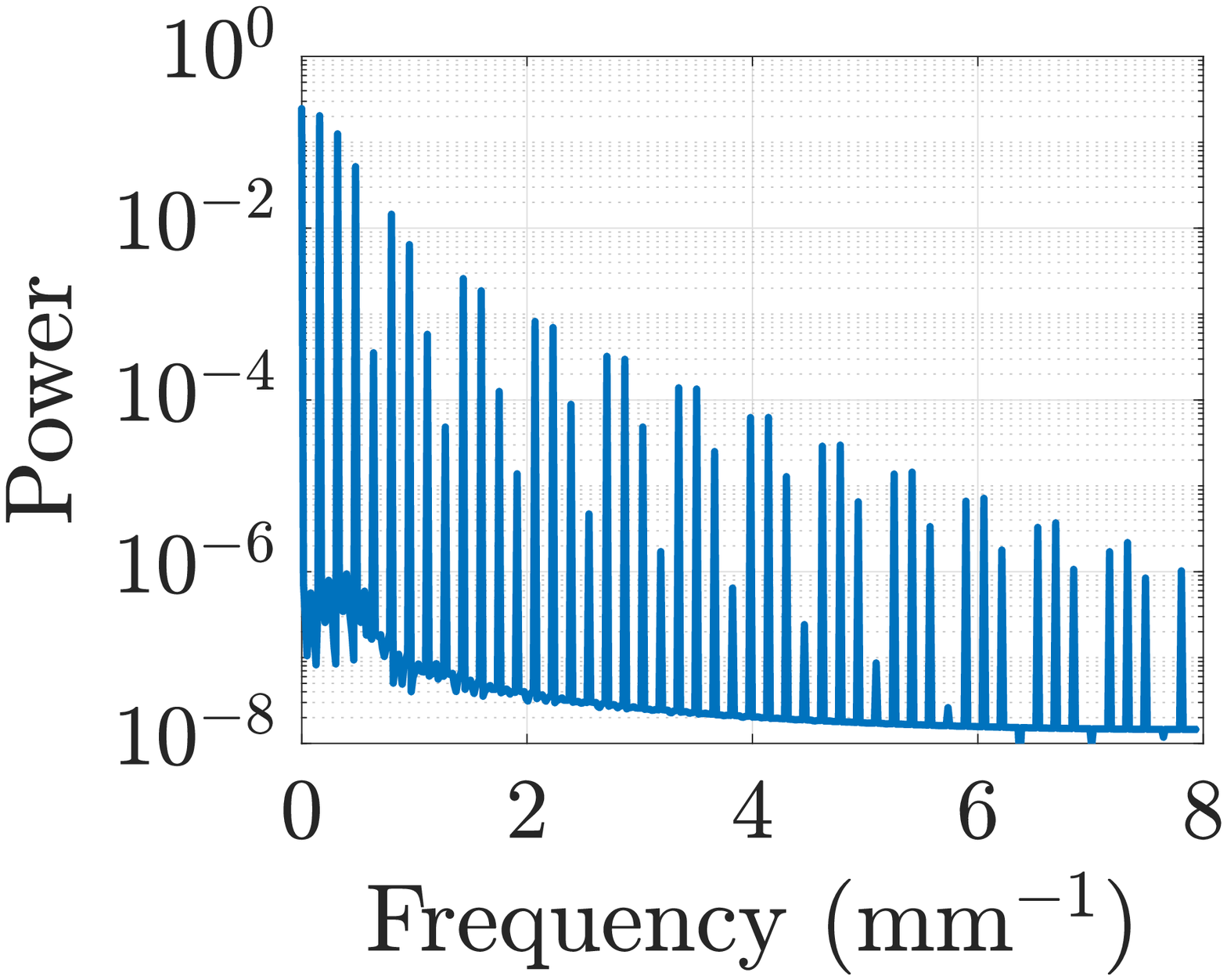}
    \end{subfigure}
    \begin{subfigure}{0.49\linewidth}
    \caption{}
    \label{fig:fundpi2powerc}
    \includegraphics[width=4.1cm]{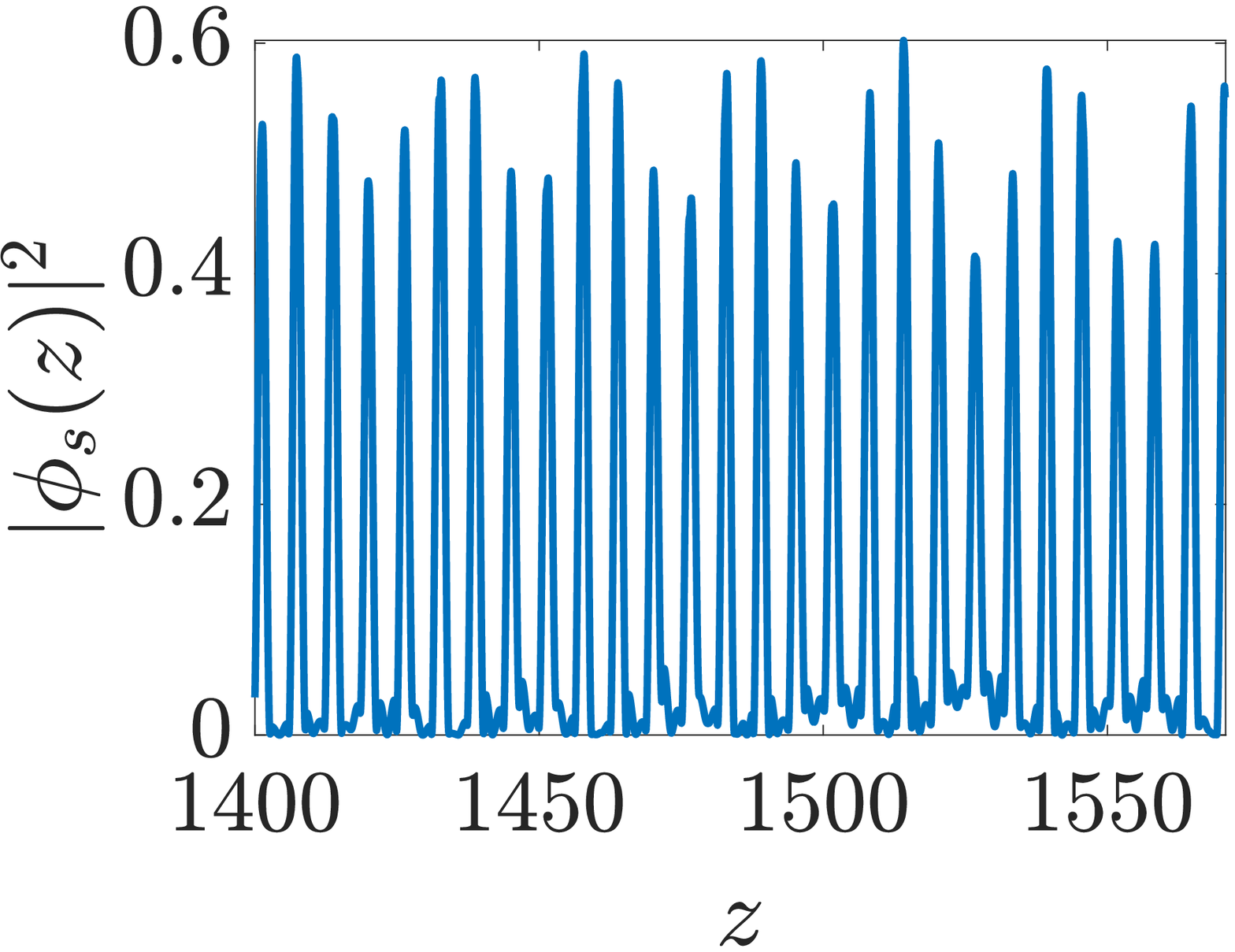}
    \end{subfigure}
    \begin{subfigure}{0.49\linewidth}
    \caption{}
    \label{fig:fundpi2powerd}
    \includegraphics[width=4.1cm]{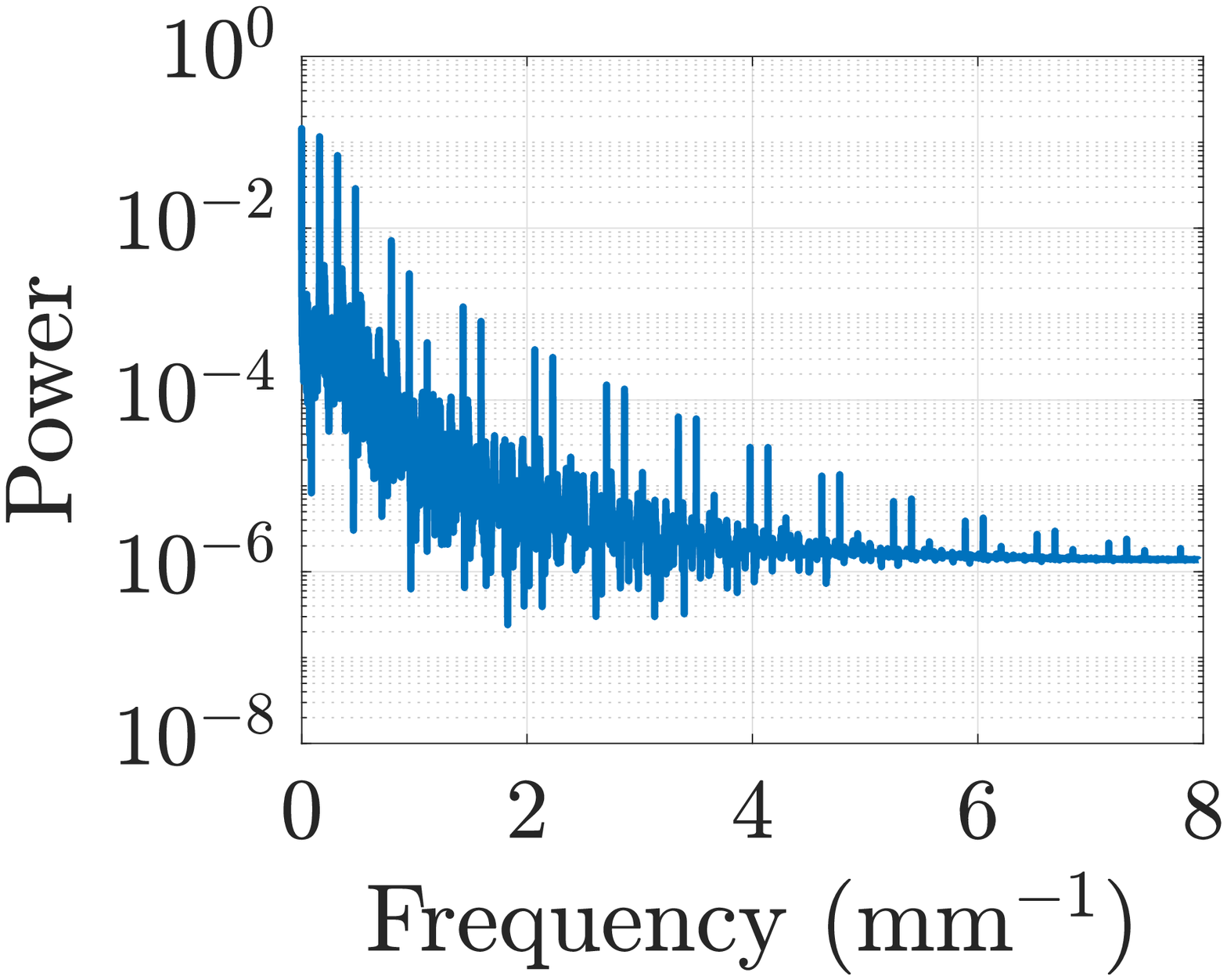}
    \end{subfigure}
    \caption{(a)-(b) Evolution in $z$ (a) and log power spectrum (b) of central site 1 (see \cref{fig:fundbreathera}) for the unperturbed fundamental breather on $z$ interval $[0,10\pi]$. (c)-(d) Evolution in $z$ (c) and log power spectrum (d) of central site 1 for the perturbed fundamental breather from \cref{fig:fundpi2pertb} on the $z$ interval $[250\pi,500\pi]$; this corresponds to the flat part of \cref{fig:fundpi2pertb}, after the perturbation growth has sufficiently saturated. Only the end of this interval is shown in (c). $20 \times 20$ lattice, $C = 1.19$, period $T = 2\pi$, phase shift $\theta = \pi/2$, steepness factor $k = 10$.}
    \label{fig:fundpi2powerspec}
\end{figure}

Next, we consider the case when the phase shift is $\theta = \pi/2$, in which
case the breather period is $\tau=8 \pi$.
When $C=1$, numerical simulations, both from spectral computations and evolution experiments, suggest that the fundamental breather solution is stable. The behavior is qualitatively the same as when $\theta = \pi$. However, when $C$ is increased from 1 by parameter continuation (\cref{fig:fundbgangle}), an unstable Floquet eigenmode appears at a critical value of $C$ (between $C=1.16$ and $C=1.17$ for our chosen parameters; see \cref{fig:fundpi2a}). This unstable eigenmode is not on the real axis, i.e. it corresponds to a complex quartet (see inset in \cref{fig:fundpi2a}, as well as \cref{fig:fundpi2b}, which shows the growth of this mode in $C$), and it does not depend on the size of the grid (in contrast with what occurs for $\theta = \pi$). 
Its spatial profile is shown in \cref{fig:fundpi2perta}. In addition, the unstable mode is not part of the continuous spectrum, in that it is not present in the linearization about the background state. Once again, we can see the consequences of this unstable Floquet eigenmode by perturbing the fundamental breather through adding a small multiple of the corresponding eigenfunction (\cref{fig:fundpi2pertb}). The slope of the least squares regression line in the figure is within 2\% of  $\log |\lambda|^{1/\tau}$, where $\lambda$ is the unstable Floquet multiplier. The log power spectrum of a single central site in the unperturbed and perturbed fundamental breather is shown in \cref{fig:fundpi2powerspec}. While in both cases the fundamental frequency is the rotation frequency $1/(2 \pi)$, it is evident that the unstable regime
leads to a genuinely distinct evolution at longer times that involves
the excitation of each node (via multiple intensity peaks)
throughout the period, as is clear from the left panels
of the figure. The right panels also show 
a discernibly distinct (and much faster in its
decay) tail of the frequency dependence of the intensity. 
A closer inspection of the relevant spectrum over a narrow frequency
range reveals the growth of sidebands which is in line with the
apparent quasi-periodic behavior observed in \cref{fig:fundpi2powerc}.

\subsection{Other phase shifts}

Numerical computations suggests that the fundamental breather solution exists for phase shifts of $\pi/3$ and $\pi/4$. However, in both of these cases, computation of the Floquet spectrum for $C=1$ shows the presence of an unstable Floquet multiplier on the real axis, hence we do not further pursue such waveforms
herein.

\section{Two-site breathers on the unit square}

\begin{figure}
    \centering
    \begin{subfigure}{0.65\linewidth}
    \caption{}
    \label{fig:breatherdiaga}
    \includegraphics[width=5.5cm]{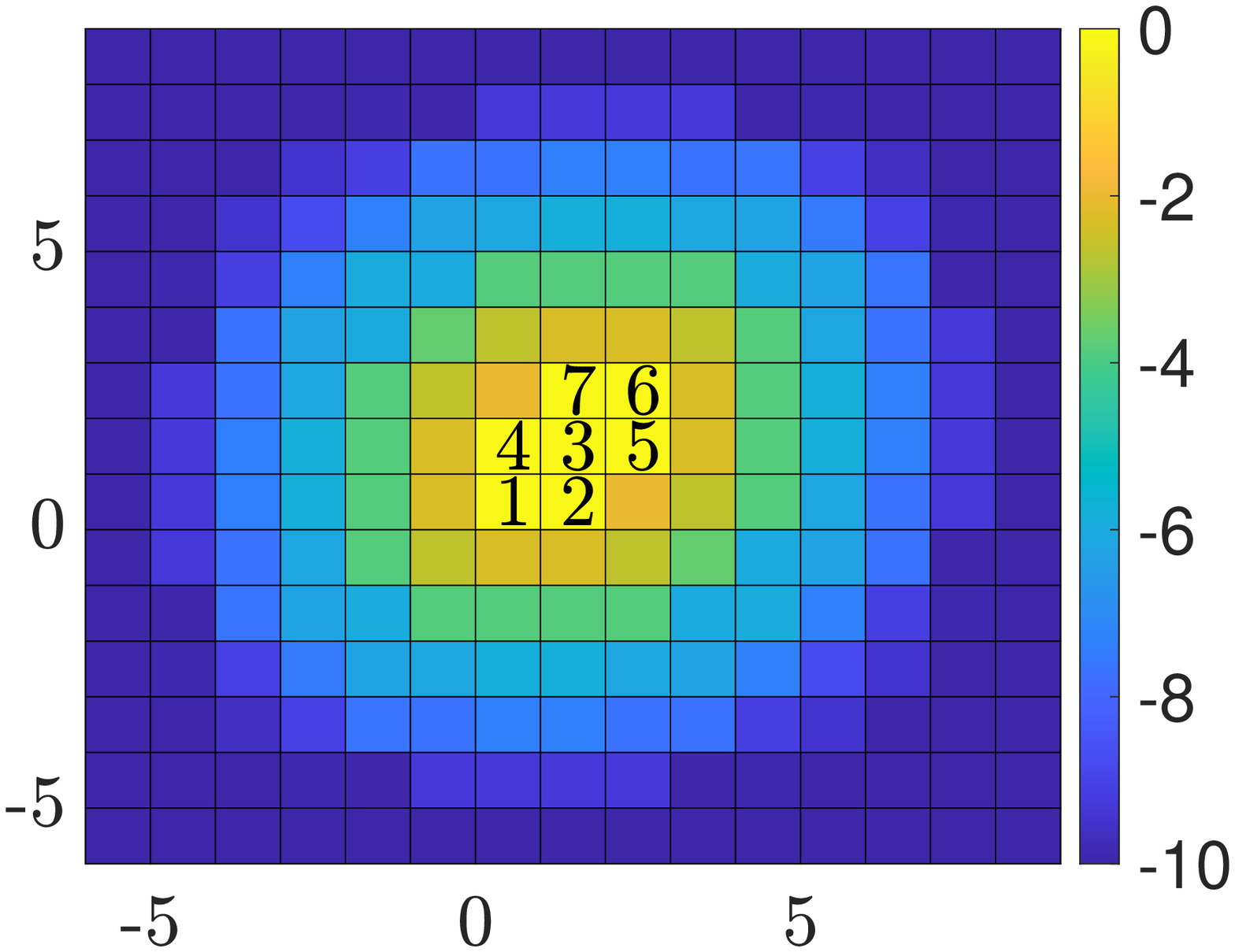}
    \end{subfigure}
    \begin{subfigure}{0.3\linewidth}
    \caption{}
    \label{fig:breatherdiagb}
    \includegraphics[width=2cm]{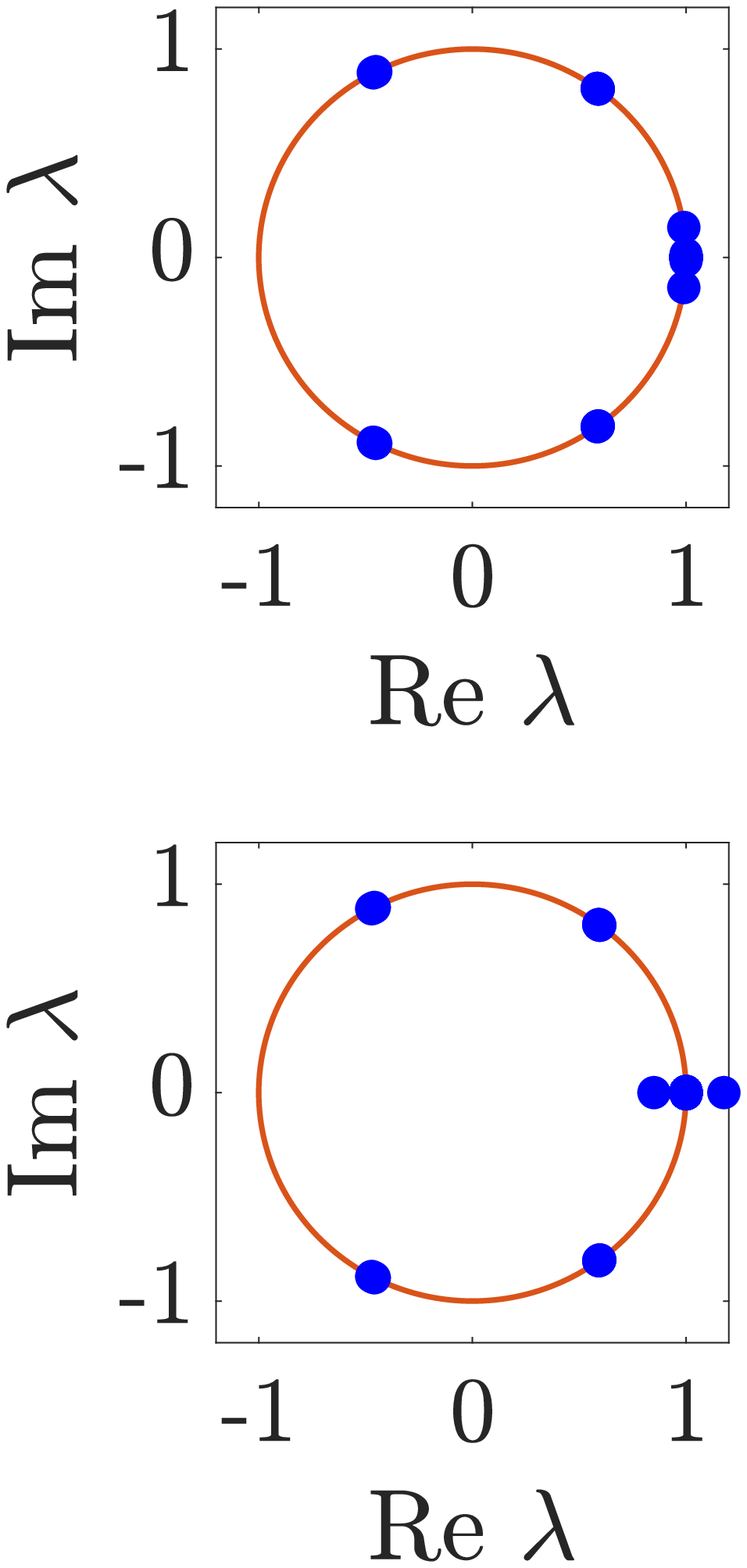} 
    \end{subfigure}
    \begin{subfigure}{0.49\linewidth}
    \caption{}
    \label{fig:breatherdiagc}
    \includegraphics[width=4.1cm]{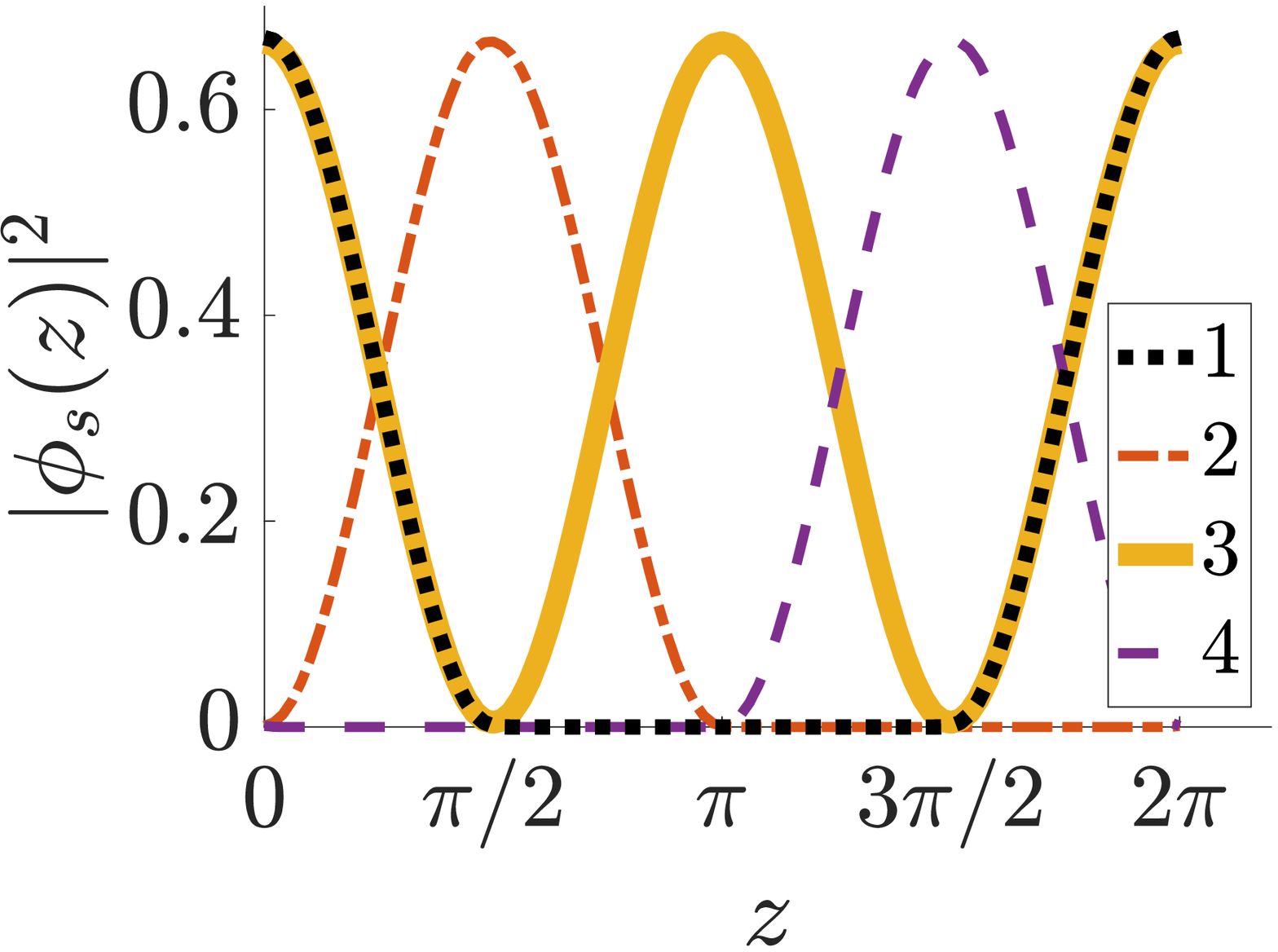}
    \end{subfigure}
    \begin{subfigure}{0.49\linewidth}
    \caption{}
    \label{fig:breatherdiagd}
    \includegraphics[width=4.1cm]{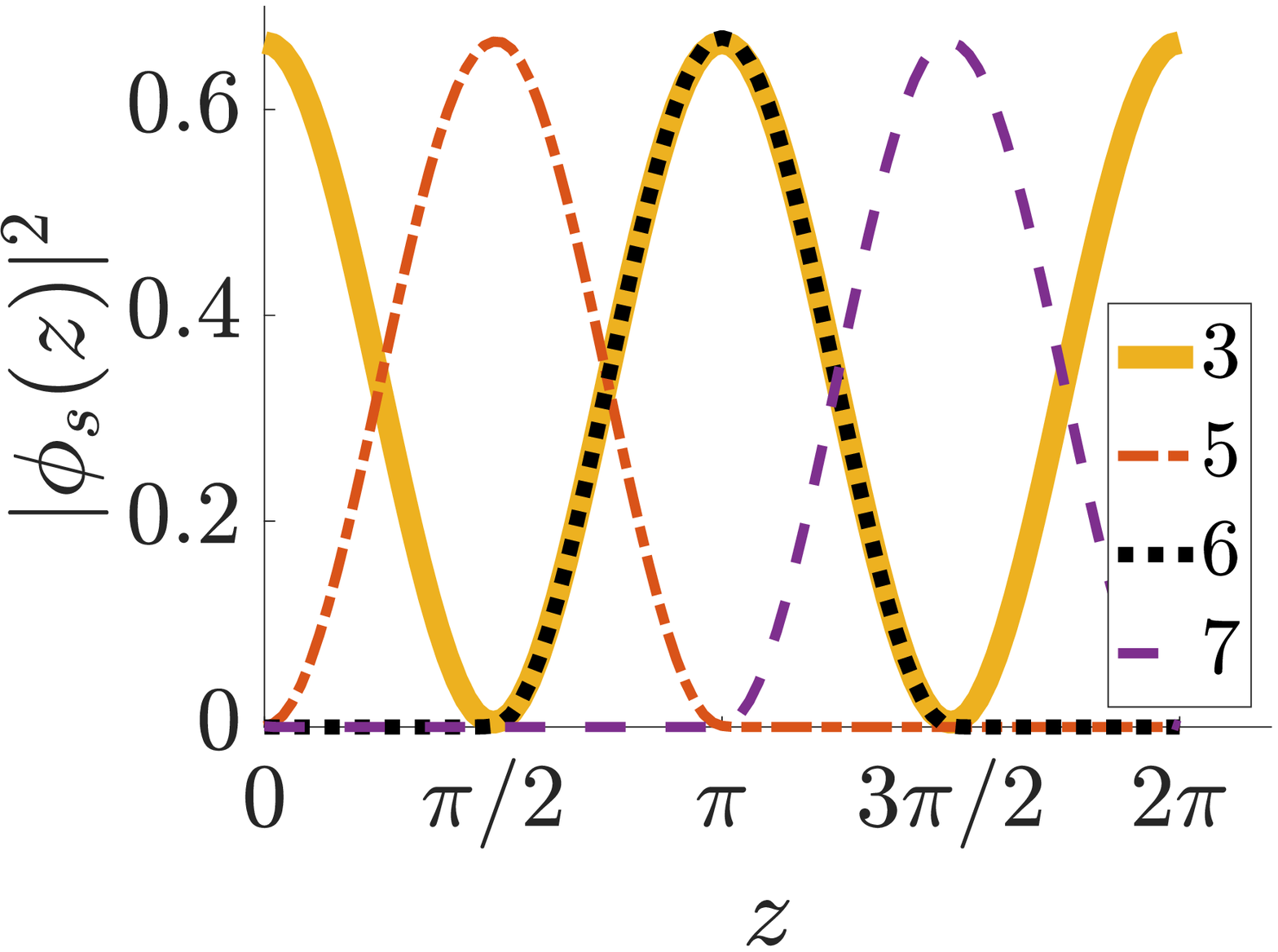}
    \end{subfigure}
    \caption{(a) Color map of the $\log_{10}$ max intensity at each site of the diagonal breather over one period. Note that the intensity is the same for the in-phase and out-of-phase breathers. (b) Floquet spectrum of diagonal breather for opposite sites initialized out-of-phase (top) and in-phase (bottom). For the unstable in-phase breather, the largest Floquet multiplier is $\lambda = 1.1770$. (c)-(d) Intensity of the solution at the primary sites where the breather is localized. Note that the intensity at site 3 partially overlaps with site 1 in (c) and with site 6 in (d). The intensity in (a), (c), and (d) is the same for in-phase and out-of-phase breathers, although their stability is not.
    $20\times 20$ lattice, coupling constant $C = 1$, period $T = 2\pi$, phase shift $\theta = \pi$, steepness factor $k = 10$.}
    \label{fig:breatherdiag}
\end{figure}

\begin{figure}
    \centering
    \begin{subfigure}{0.65\linewidth}
    \caption{}
    \label{fig:breatheradja}
    \includegraphics[width=5.5cm]{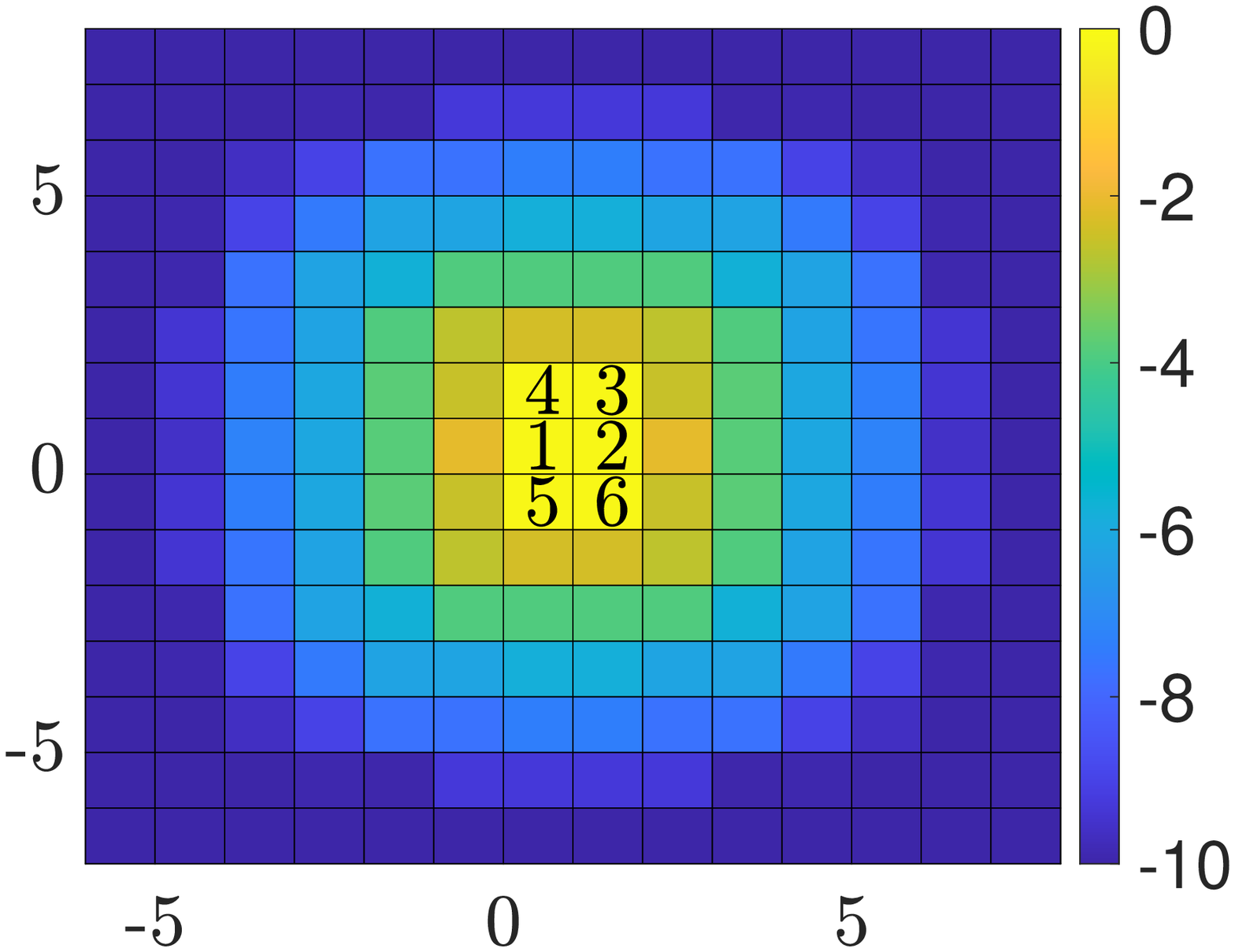}
    \end{subfigure}
    \begin{subfigure}{0.3\linewidth}
    \caption{}
    \label{fig:breatheradjb}
    \includegraphics[width=2cm]{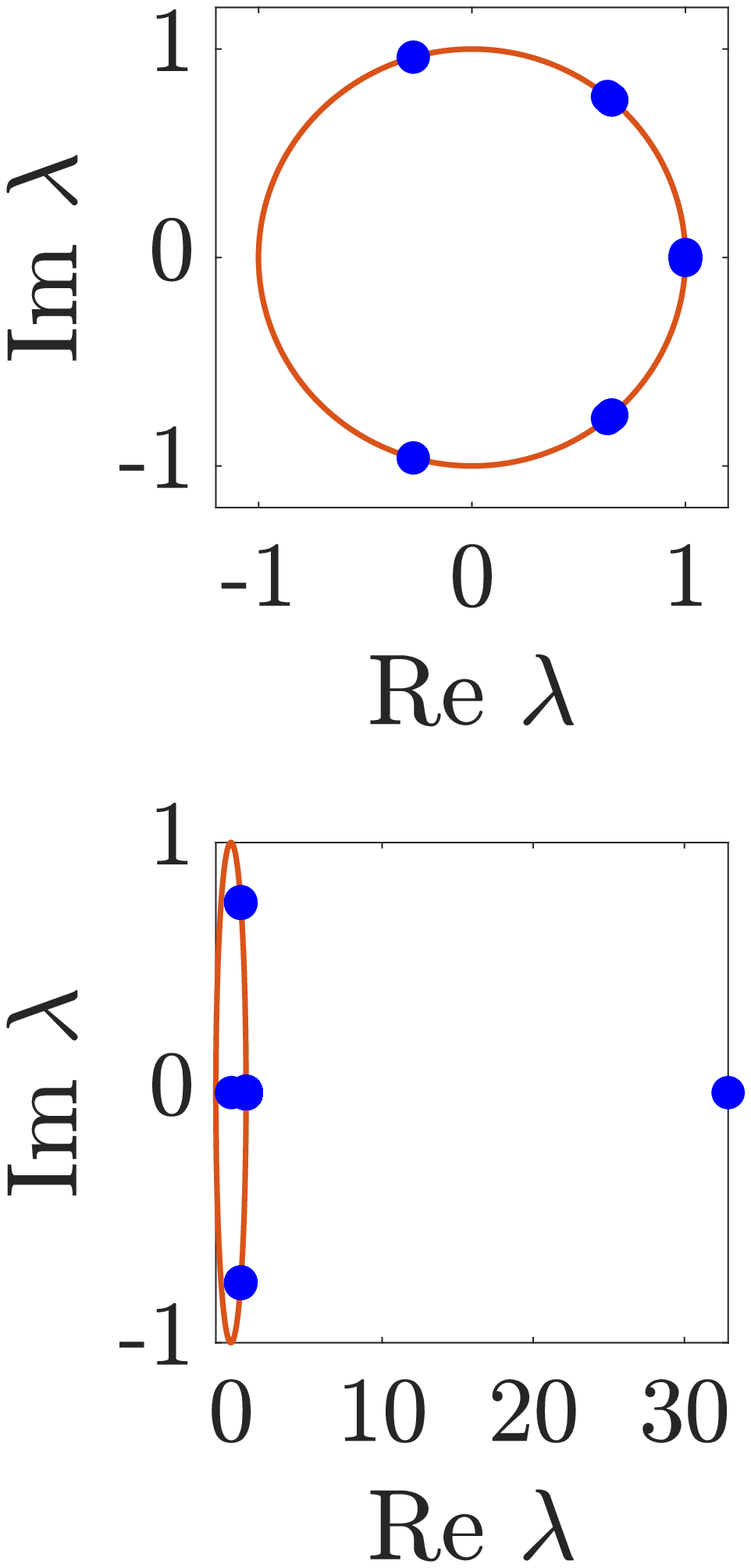} 
    \end{subfigure}
    \begin{subfigure}{0.49\linewidth}
    \caption{}
    \label{fig:breatheradjc}
    \includegraphics[width=4cm]{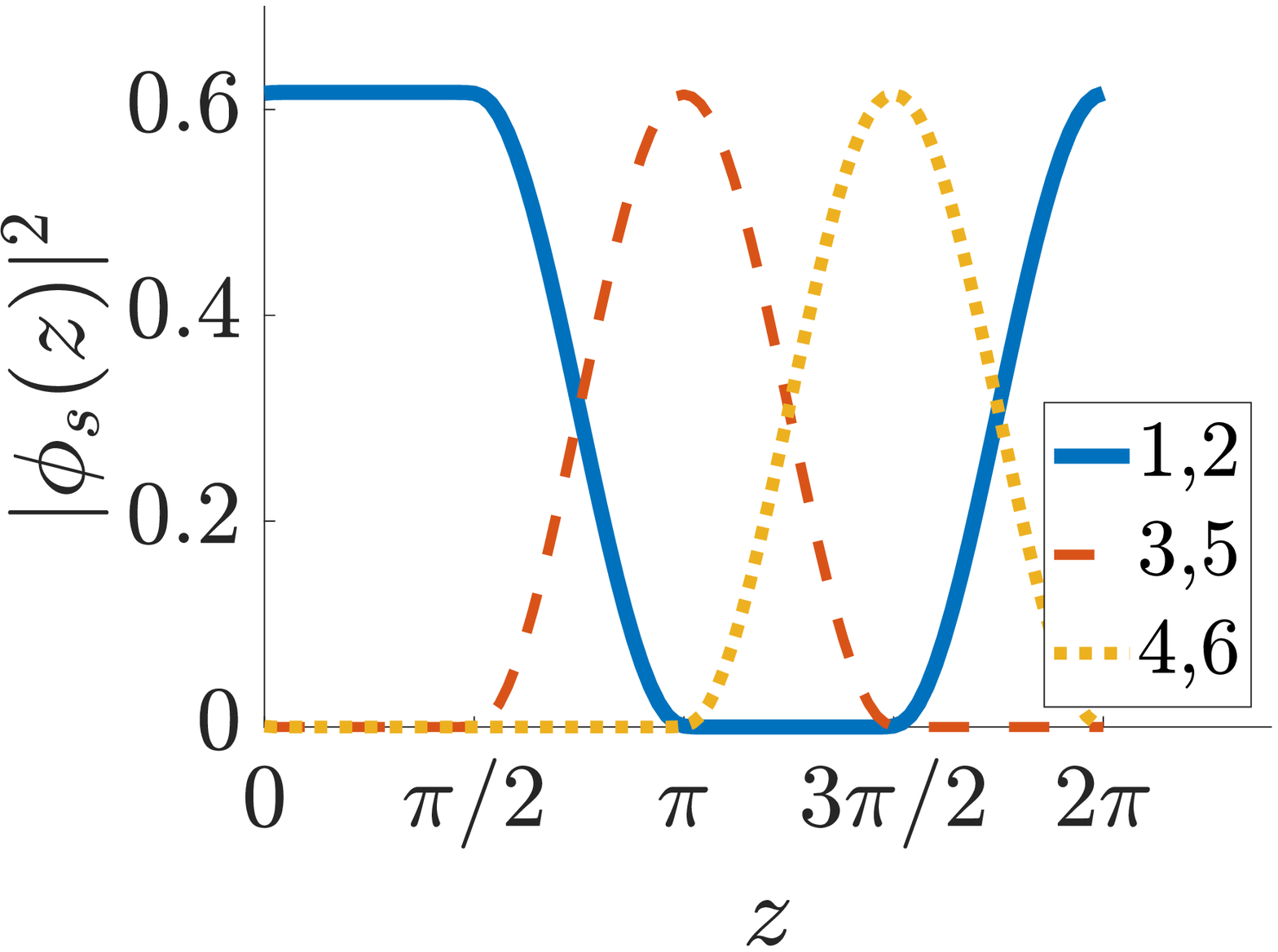}
    \end{subfigure}
    \begin{subfigure}{0.49\linewidth}
    \caption{}
    \label{fig:breatheradjd}
    \includegraphics[width=4cm]{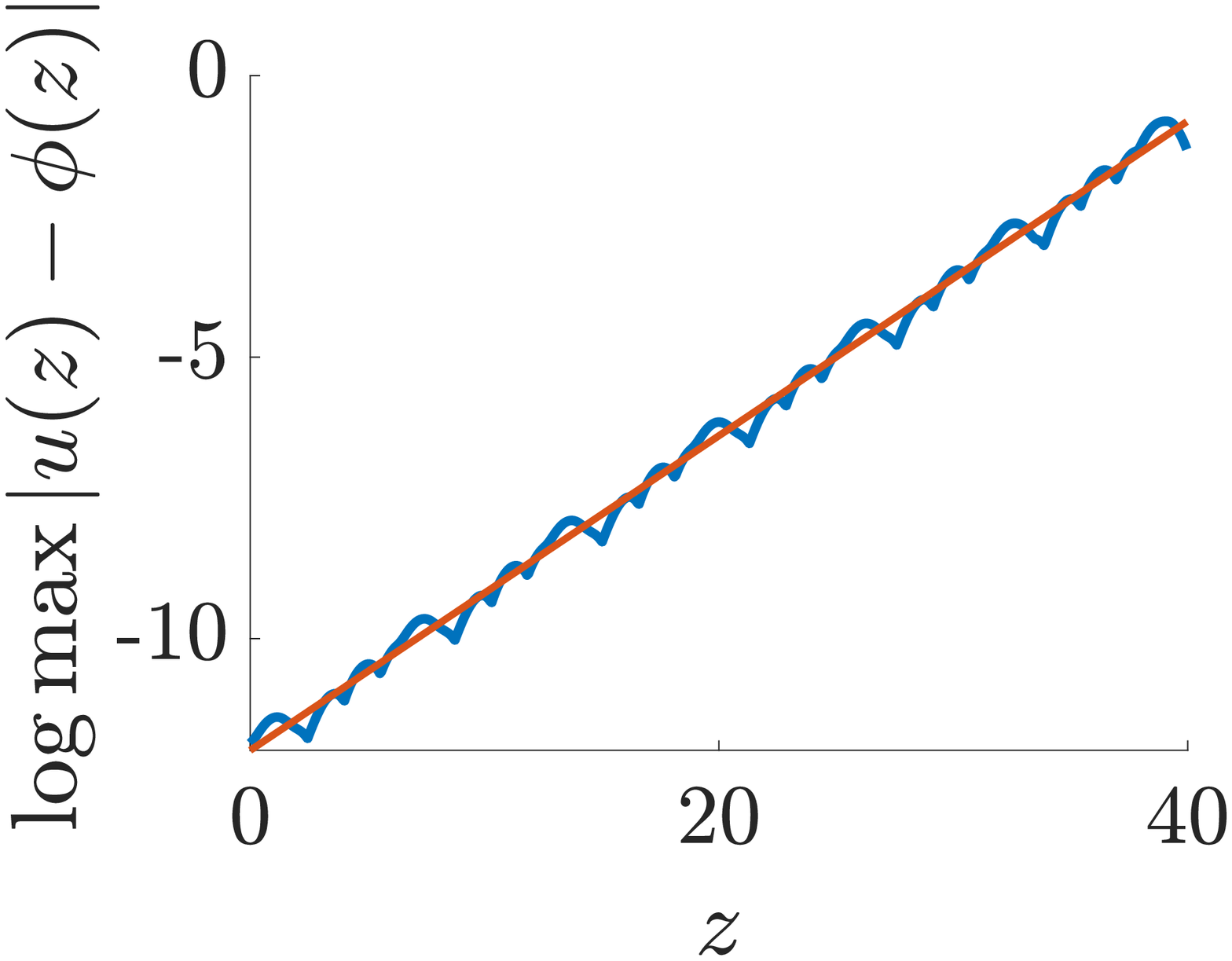}
    \end{subfigure}
    \caption{(a) Color map of the $\log_{10}$ max intensity at each site of the adjacent breather over one period. (b) Floquet spectrum of the adjacent breather for adjacent sites initialized out-of-phase (top) and in-phase (bottom). For the unstable in-phase breather, the largest Floquet multiplier is $\lambda = 32.8933$. (c) Intensity of the solution at the primary sites where the breather is localized. The intensity in (a) and (c) is the same for in-phase and out-of-phase breathers, although their stability is not. (d) Log of the maximum absolute difference between the perturbed solution $u(z)$ and the in-phase adjacent breather $\phi(z)$; initial condition for perturbation is $\phi(0)+\epsilon v$, where $\epsilon = 10^{-5}$ and $v$ is the eigenfunction corresponding to the largest Floquet multiplier. The least-squares linear regression line is also shown with slope 0.2783. $20\times 20$ lattice, coupling constant $C = 1$, period $T = 2\pi$, phase shift $\theta = \pi$, steepness factor $k = 10$.}
    \label{fig:breatheradj}
\end{figure}

Multi-breather solutions can be found for which the initial intensity is localized at more than one site in the lattice. 
Here, in line with the discussion in the Supplemental Material
of~\cite{Mukherjee2020}, we envision a
scenario whereby light is launched initially on two waveguides
rather than a single one.
We will consider here two-site breathers, in which the initial intensity is localized at a pair of sites within the fundamental unit square. For all of these solutions, we will take $\theta = \pi$. There are two possibilities for these two-site breathers, in terms of location: adjacent (sites 1 and 2 in \cref{fig:fundbreather}) and diagonal (sites 1 and 3 in \cref{fig:fundbreather}). In addition, for each of the two-site breather possibilities, there are two scenarios
in terms of the relative phase between the sites. If the two sites are initialized in-phase, there is a Floquet eigenvalue outside the unit circle, thus this solution is unstable. If the two sites are initialized out-of-phase, the Floquet eigenvalues all lie on the unit circle (for $C$ close to 1), which suggests that this solution is stable. (See figures \cref{fig:breatherdiagb} and \cref{fig:breatheradjb}). Of course, as 
is well documented for their DNLS analogues~\cite{kev09}, 
as $C$ is increased such structures can eventually
become unstable through complex instabilities due to multiplier quartets. 

We consider the diagonal breather first, in which the initial intensity is localized at two diagonally opposite sites in the fundamental unit square (sites 1 and 3 in \cref{fig:breatherdiag}). Due to the nature of the coupling, since each is only connected to a single neighboring site for a given $z$, the two components of the breather practically act independently. Over one period, the initial intensity at site 1 rotates counterclockwise around the fundamental unit square (sites 1,2,3,4 in \cref{fig:breatherdiag}). Since the edge between sites 3 and 5 is also active at $z= 0$ (see \cref{fig:model}), the initial intensity at site 3 rotates counterclockwise around the unit square located one position to the northeast (sites 3,5,6,7 in \cref{fig:breatherdiag}). Since site 3 is shared between both components of the multi-breather, its frequency is twice that of the other sites (\cref{fig:breatherdiagc} and \cref{fig:breatherdiagd}).

For the adjacent breather, the initial intensity is localized at two adjacent sites in the fundamental unit square (sites 1 and 2 in \cref{fig:breatheradj}). The initial intensity at site 1 rotates counterclockwise around the fundamental unit square as before (sites 1,2,3,4 in \cref{fig:breatheradj}). Since the only active connection involving site 2 at $z=0$ is that between sites 2 and 1 (see \cref{fig:model}), the initial intensity at site 2 also rotates counterlockwise around the unit square located one position to the south (sites 2,1,5,6 in \cref{fig:breatheradj}). This means that sites 1 and 2 are active for the first half of the period (\cref{fig:breatheradjc}).

\begin{figure}
    \centering
    \begin{subfigure}{0.49\linewidth}
    \caption{}
    \label{fig:breatherunstableeiga}
    \includegraphics[width=4cm]{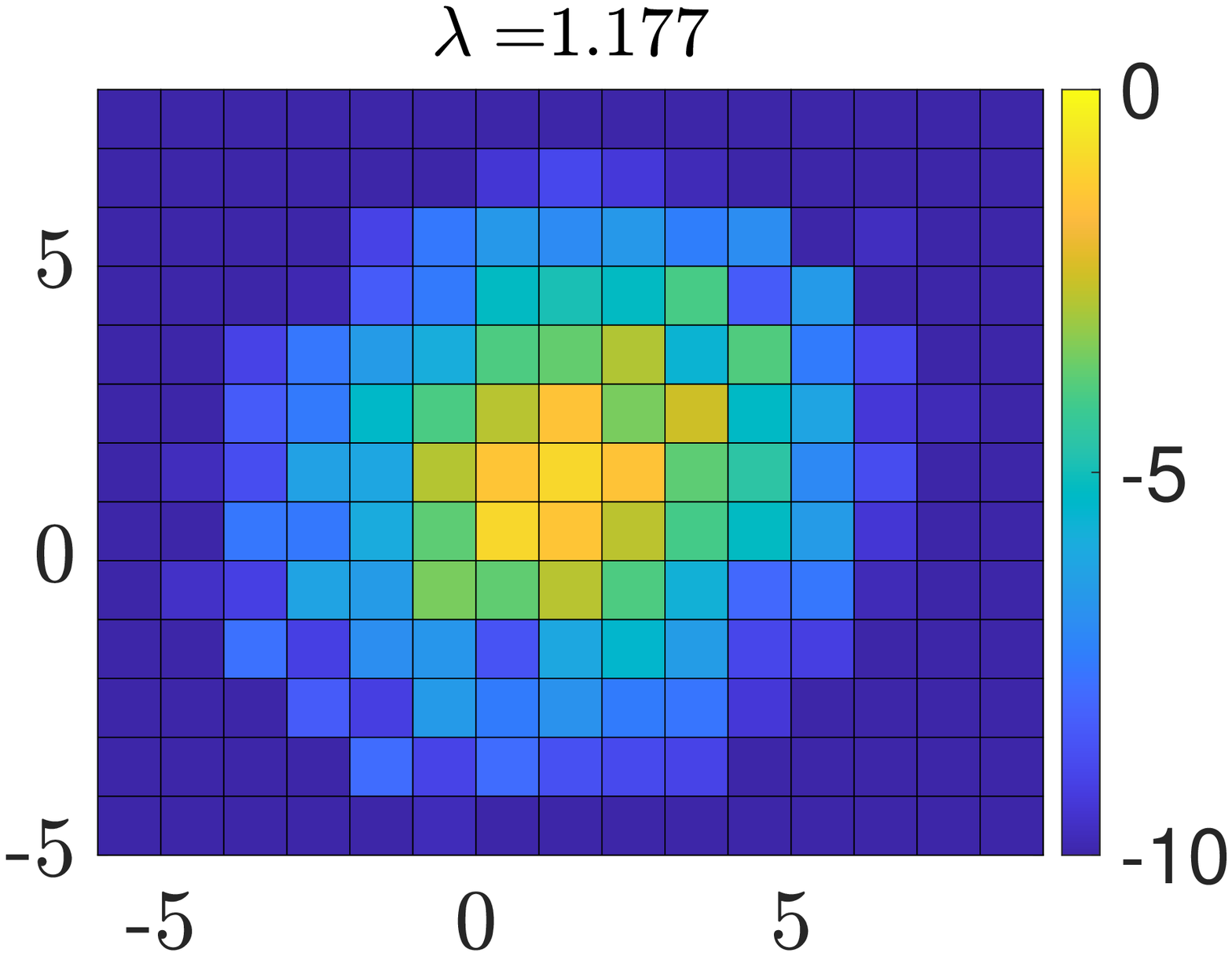}
    \end{subfigure}
    \begin{subfigure}{0.49\linewidth}
    \caption{}
    \label{fig:breatherunstableeigb}
    \includegraphics[width=4cm]{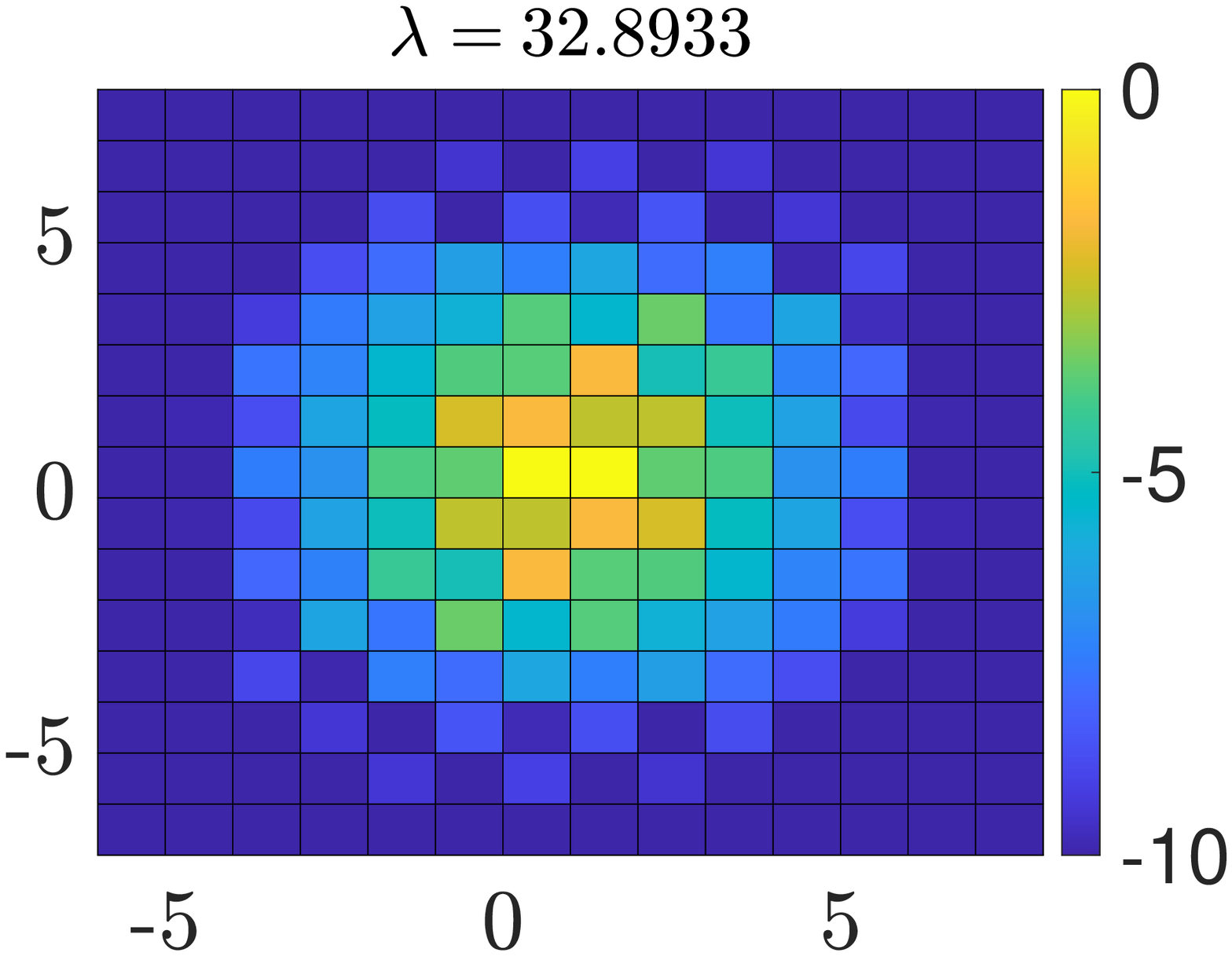}
    \end{subfigure}
    \caption{(a)-(b) Color map of the $\log_{10}$ max intensity at each site for the Floquet eigenfunction corresponding to the largest Floquet multiplier for the unstable diagonal breather (a) and the unstable adjacent breather (b).
    One can note the different spatial structure of the corresponding
    eigenfunctions. $20\times 20$ lattice, coupling constant $C = 1$, period $T = 2\pi$, phase shift $\theta = \pi$, steepness factor $k = 10$.}
    \label{fig:breatherunstableeig}
\end{figure}

In both cases, if the two sites are initialized in-phase, there is a Floquet eigenvalue outside the unit circle; this eigenvalue is much larger for the adjacent breather than for the diagonal breather. If the two adjacent sites are initialized with opposite phases, the Floquet spectrum lies on the the unit circle (for $C$ close to 1). The Floquet eigenfunctions corresponding to the largest Floquet multiplier for both the unstable diagonal breather and unstable adjacent breather are shown in \cref{fig:breatherunstableeig}.

For the unstable adjacent two-site breathers, we can see how perturbations evolve by adding a small amount of the unstable Floquet eigenfunction to the initial condition (see \cref{fig:breatheradjd}). The slope of the least squares linear regression line is 0.2783, which is a relative error of less than $10^{-3}$ from the predicted value of $\log |\lambda|^{1/\tau}$, where $\lambda = 32.8933$ is the value of the largest Floquet multiplier, and $\tau = 4 \pi$ is the period of the breather. Similar results can be obtained for the unstable opposite two-site breather.

\begin{figure}
    \centering
    \begin{subfigure}{0.49\linewidth}
    \caption{}
    \label{fig:diagoppperta}
    \includegraphics[width=4cm]{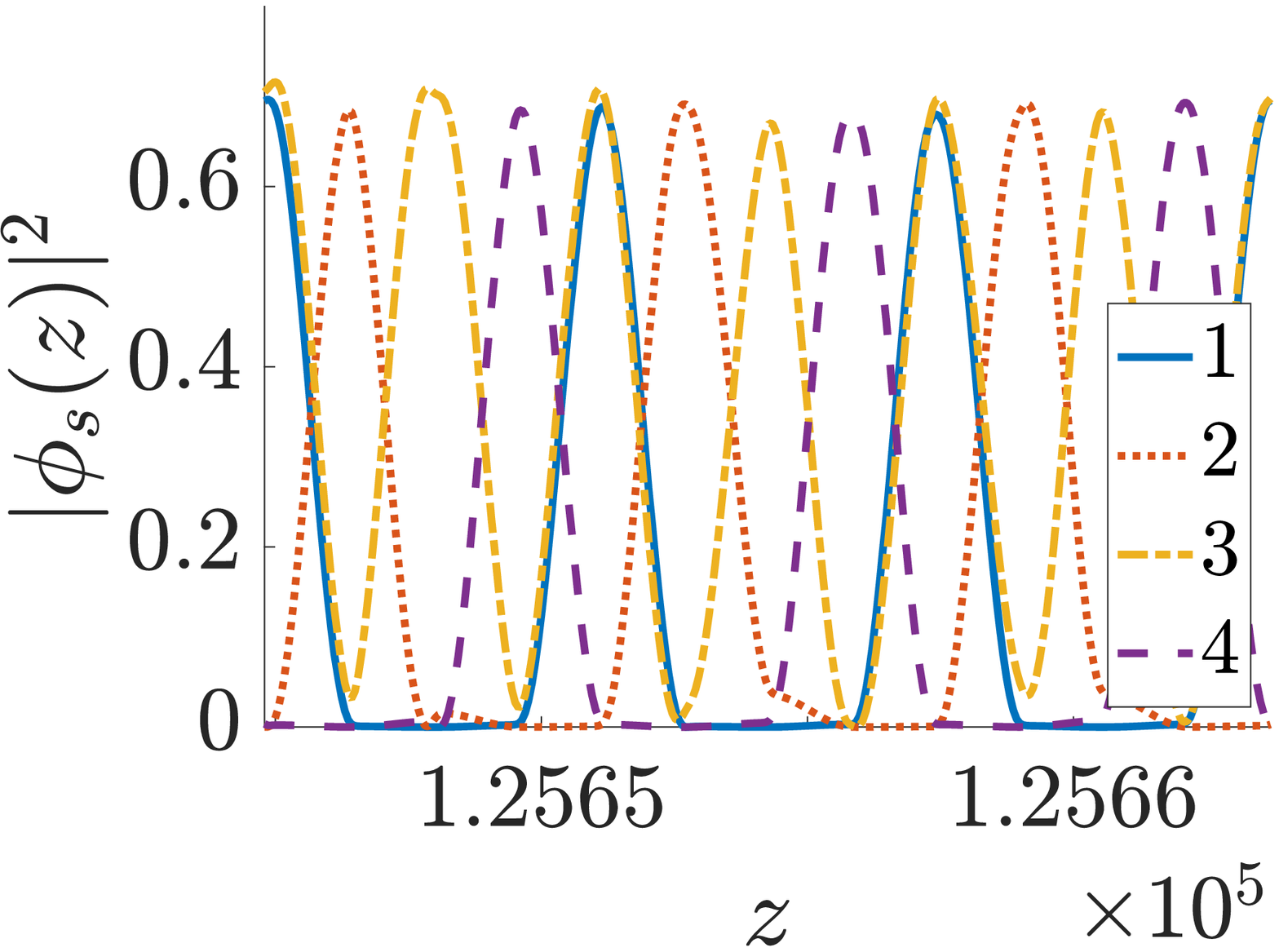}
    \end{subfigure}
    \begin{subfigure}{0.49\linewidth}
    \caption{}
    \label{fig:diagopppertb}
    \includegraphics[width=4cm]{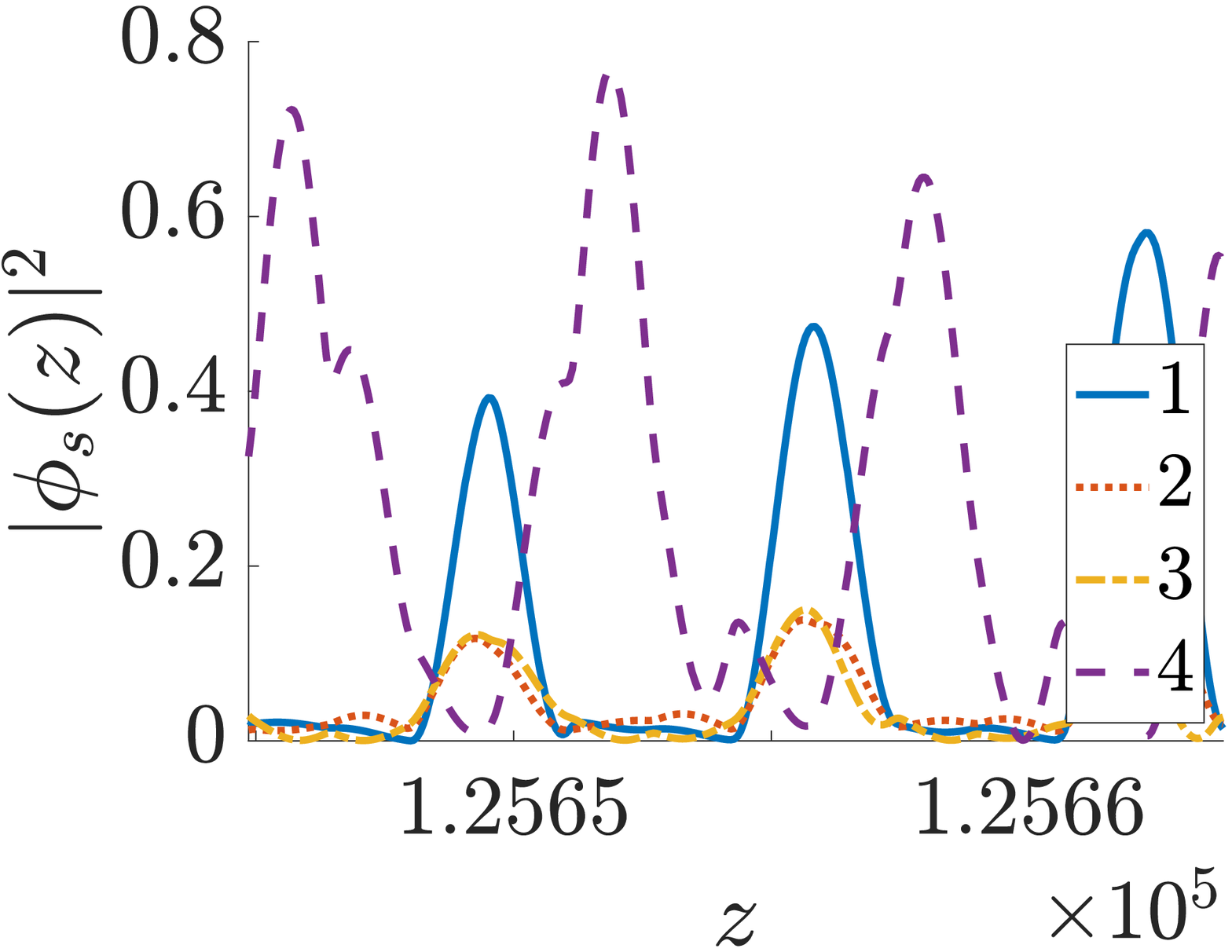}
    \end{subfigure}
    \caption{(a)-(b) Solution for four of the seven central sites at the end of the $z$ interval for the evolution of the perturbed opposite breather with out-of-phase initialization (a) and in-phase initialization (b). Initial condition is two adjacent sites initialized with the same intensity. $20\times 20$ lattice, coupling constant $C = 1$, period $T = 2\pi$, phase shift $\theta = \pi$, steepness factor $k = 10$.}
    \label{fig:diagopppert}
\end{figure}

\begin{figure}
    \centering
    \begin{subfigure}{0.49\linewidth}
    \caption{}
    \label{fig:breatheradjtimestepa}
    \includegraphics[width=4cm]{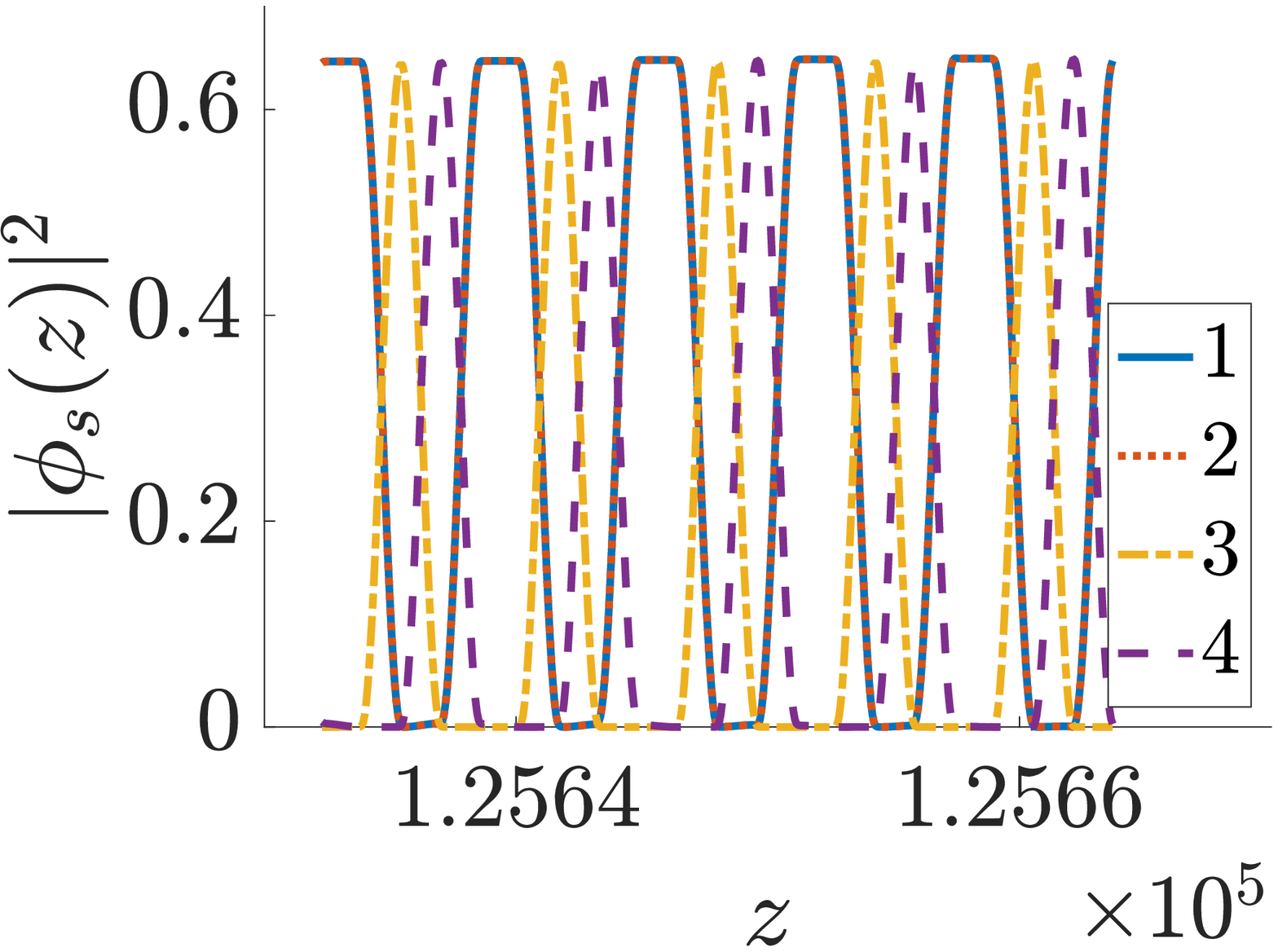} 
    \end{subfigure}
    \begin{subfigure}{0.49\linewidth}
    \caption{}
    \label{fig:breatheradjtimestepb}
    \includegraphics[width=4cm]{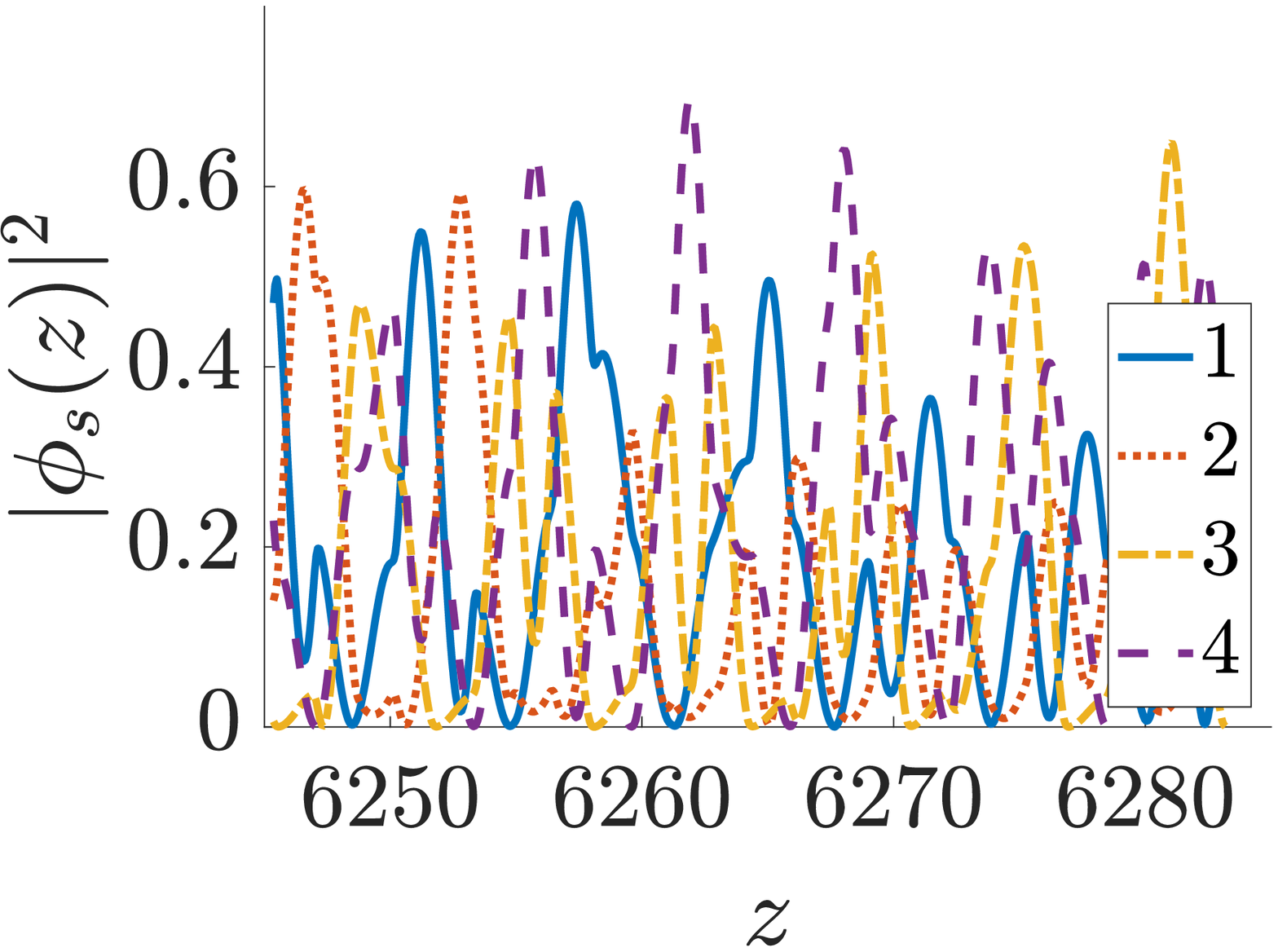}
    \end{subfigure}
    \caption{(a)-(b) Solution for four of the six central sites at the end of the $z$ interval for the evolution of the perturbed adjacent breather with out-of-phase initialization on $[0, 10^5 \pi]$ (a) and the unperturbed adjacent breather with in-phase initialization on $[0, 2000 \pi]$ (b). Initial condition is two adjacent sites initialized with the same intensity. 
    $20\times 20$ lattice, coupling constant $C = 1$, period $T = 2\pi$, phase shift $\theta = \pi$, steepness factor $k = 10$.}
    \label{fig:breatheradjtimestep}
\end{figure}

Long-term evolution numerical experiments confirm these results, and provide evidence that the diagonal breather with opposite sites initialized out-of-phase is dynamically robust (for the parameter range and propagation distance considered herein). For the initial condition, we take two diagonally opposite sites in the unit square which have the same intensities, but are out of phase by $\pi$. This initial amplitude is chosen to be the maximum amplitude of the diagonal breather. Results of this evolution are shown in \cref{fig:diagoppperta}. Similarly, the diagonal breather with opposite sites initialized in-phase is unstable, although it takes many steps for a perturbation of this solution to break apart and lead to
a distinct nearly periodic orbit, as indicated in \cref{fig:diagopppertb}. Similarly, numerical evolution experiments for the adjacent breather confirm that it is stable for the out-of-phase configuration (\cref{fig:breatheradjtimestepa}) and unstable for the in-phase configuration (\cref{fig:breatheradjtimestepb}). In fact, the instability of the adjacent breather with in-phase initialization manifests itself in a way such that it breaks apart by $z=2000 \pi$.
% Comparison of the power spectrum of this solution before and after it breaks apart (\cref{fig:breatheradjtimestepc} and \cref{fig:breatheradjtimestepd}) suggests the solution does not settle into a clearly discernible pattern for the $z$ interval of our numerical computations.

\section{Conclusions \& Future Challenges}

In the present work, we have explored a wide set of Floquet solitary wave structures in a system bearing a topological
bandgap. In particular, we were motivated by direct numerical simulations
and experimental observations in a photonic implementation of
system with waveguides bearing a time-modulated coupling
structure in two spatial dimensions. We were able to identify
prototypical time-periodic solutions of the system in the 
form of fundamental breathers bearing different phase shifts
upon completion of a period of the time variation of the coupling.
We analyzed the Floquet spectrum of such solitons, 
distinguishing their continuous spectrum (and its dependence on the coupling parameter),
as well as the point spectrum associated with the excited sites
of the relevant coherent structure. We found that, aside from
lattice-size-dependent oscillatory instabilities, the 
fundamental breathers were spectrally, as well as dynamically robust.
We then moved one step further, exploring multi-peak (excited)
structures. We leveraged a detailed understanding of the spectral
picture of such structures in the stationary DNLS limit to explain
the corresponding stability analysis of excited, multi-peaked
time-periodic states. In particular, we found that double-peaked states are unstable 
if the two sites are initialized in-phase, and spectrally stable if the two
sites are initialized out-of-phase. This relationship between phase and stability 
is similar to what is observed for multi-peaked standing wave solutions in DNLS lattices, both in one and two dimensions \cite{Kalosakas2006,Parker2020}, as well as breather solutions
in Klein-Gordon lattices \cite{Pelinovsky2012}.

Naturally, this is not a full outcome in this ongoing effort
to explore the existence, stability and dynamical properties
of topological solitonic structures. For instance, one can
consider different types of lattices, including Lieb and 
Kagom{\'e} ones, and further explore the wave patterns that arise
therein and their corresponding spectra. Another aspect 
in which topological features may have a strong imprint
is the mobility of nonlinear modes. Indeed, it
has been argued in recent works, including~\cite{Abl19a,Abl21a,recht21}, 
that topology may control and, indeed, even enhance (when suitably
leveraged) the mobility of states that might not be otherwise
particularly mobile (e.g., due to Peierls-Nabarro and associated
barriers~\cite{kev09,Abl21a}) in conventional discrete settings. It is intriguing to consider if mobility of photonic modes can be achieved in a similar way to what is seen in the propagation of nonlinear elastic waves in flexible structures which provides opportunities for locomotion of mechanical robots ~\cite{Bertoldi2}. 
Having focused herein on stationary states, 
such features are worthwhile of further exploration and we
defer corresponding studies to future publications.

\begin{acknowledgments}
This material is based upon work supported by the U.S. National Science Foundation under the RTG grant DMS-1840260 (R.P. and A.A.), DMS-1809074 (P.G.K.), and DMS-1909559 (A.A.). J.C.-M. acknowledges support from EU (FEDER program 2014-2020) through both Consejería de Economía, Conocimiento, Empresas y Universidad de la Junta de Andalucía (under the projects P18-RT-3480 and US-1380977), and MICINN and AEI (under the projects PID2019-110430GB-C21 and PID2020-112620GB-I00).
\end{acknowledgments}

\appendix

\section{Continuous spectrum bands}\label{app:bands}

We compute the continuous spectrum bands in the limiting case where the coupling functions $J_j(z)$ are step functions. We note that although these step functions are not everywhere differentiable, we can treat the $z$-dependent term $H_{ss'}$ in \cref{eq:model} as piecewise constant. First, we linearize equation \cref{eq:model} about the solution $\phi_s = 0$, which is equivalent to only considering the linear terms. We then take the Floquet ansatz $\phi_s(z) = e^{\lambda z}u_s(z)$, where $u_s(z)$ is periodic in $z$ with period $\tau = m T$, and $m = \frac{2 \pi}{\theta}$. Substituting this into \cref{eq:model} and simplifying, we obtain the sequence of equations 
\begin{equation}\label{eq:linmodel}
\begin{aligned}
\frac{d}{dz} u_s(z) &= -\left( i C A_j + \lambda \right)u_s(z) && z \in \left[\frac{(j-1) T}{4}, \frac{jT}{4} \right)
\end{aligned}
\end{equation}
for $j = 1, \dots, 4$, which is extended periodically for all $z$. The $A_j$ are piecewise constant linear operators which implement the lattice couplings in \cref{fig:modelcartoon}. When $C=0$, the  only solutions with period $\tau$ are when $\lambda \tau$ is an integer multiple of $2 \pi$, which implies that there is a single Floquet multiplier at $\mu=1$ with infinite multiplicity.
Over one period $\tau = m T$, the solution to \cref{eq:linmodel} is given by
\begin{equation}\label{eq:linear1period}
u_s(\tau) = e^{-4 m \lambda \frac{T}{4}} B^m(C) u_s(0),
\end{equation}
where
\begin{equation}
B(C) = e^{-i C A_4 \frac{T}{4}}e^{-i C A_3 \frac{T}{4}}e^{-i C A_2 \frac{T}{4}}e^{-i C A_1 \frac{T}{4}}.
\end{equation}
We note that the exponentials in this product do not commute, since the operators $A_j$ do not commute. For $\lambda$ to be in the continuous spectrum, we require $u_s(\tau) = u_s(0)$, i.e. the operator on the RHS of \cref{eq:linear1period} must be the identity. This is equivalent to
\begin{equation}\label{eq:lineareigs1}
B^m(C) = e^{\lambda \tau }, 
\end{equation}
thus the Floquet multipliers $\mu$ are exactly the eigenvalues of $B^m$. 

We can compute these by approximating the integer lattice $\Z^2$ with successively larger finite lattices. For a square lattice of size $2N \times 2N$, with periodic boundary conditions imposed on the couplings, the operators $A_j$ are represented by $4N^2 \times 4N^2$ symmetric adjacency matrices. Computing the eigenvalues of $B(C)^m$ numerically, we verify properties (i)-(iii) of the continuous spectrum in \cref{sec:phasepi}. These results do not depend on how the lattice points are arranged in the adjacency matrices $A_j$.

\bibliographystyle{apsrev4-2}
\bibliography{references.bib}

\end{document}